%% file: main.tex
\documentclass[10pt,journal,compsoc]{IEEEtran}
\usepackage[table]{xcolor}
\usepackage{etex}
\usepackage{booktabs}
\usepackage[utf8]{inputenc}
\usepackage[ruled,vlined]{algorithm2e}
\usepackage[T1]{fontenc}
\usepackage{microtype}
\usepackage{graphicx}
\usepackage{paralist}
\usepackage{tabularx}
\usepackage{balance}
\usepackage{multicol}
\usepackage{multirow}
\usepackage{pbox}
\usepackage{enumitem}
\usepackage{amssymb}
\usepackage{pifont}
\usepackage{xspace}
\usepackage{url}
\usepackage{tikz}
\usepackage{rotating}
\usepackage{float}
\usepackage[TABBOTCAP]{subfigure}
\usepackage{ragged2e}

\ifCLASSOPTIONcompsoc
  \usepackage[nocompress]{cite}
\else
  \usepackage{cite}
\fi

  {\list{}{\leftmargin=0.1in\rightmargin=0.1in}\item[]}%
  {\endlist}

\newcommand{\ie}{\emph{i.e.,}\xspace}
\newcommand{\eg}{\emph{e.g.,}\xspace}
\newcommand{\etc}{etc.\xspace}
\newcommand{\etal}{\emph{et~al.}\xspace}
\newcommand{\secref}[1]{Section~\ref{#1}\xspace}

\newcommand{\figref}[1]{Fig.~\ref{#1}\xspace}

\newcommand{\tabref}[1]{Table~\ref{#1}\xspace}
\newcommand{\roberta}{{RoBERTa}\xspace}

\newcommand{\ngram}{{$n$-gram}\xspace}
\newcommand{\mask}{{\tt <MASK>}\xspace}

\newboolean{showcomments}

\setboolean{showcomments}{true}

\ifthenelse{\boolean{showcomments}}
  {\newcommand{\nb}[2]{
    \fbox{\bfseries\sffamily\scriptsize#1}
    {\sf\small$\blacktriangleright$\textit{#2}$\blacktriangleleft$}
   }
  }
  {\newcommand{\nb}[2]{}
  }

\begin{document}

\title{An Empirical Study on the Usage of Transformer Models for Code Completion}

\author{Matteo~Ciniselli,
        Nathan~Cooper,
        Luca~Pascarella,
        Antonio~Mastropaolo,
        Emad~Aghajani,\\
        Denys~Poshyvanyk,
        Massimiliano~Di Penta,
        and~Gabriele~Bavota% <-this % stops a space
\IEEEcompsocitemizethanks{\IEEEcompsocthanksitem M. Ciniselli is with SEART @  Software Institute, Universit\`a della Svizzera italiana, Switzerland. \protect\\
E-mail: matteo.ciniselli@usi.ch
\IEEEcompsocthanksitem N. Cooper is with SEMERU @ William \& Mary, USA. \protect\\
E-mail: nacooper01@email.wm.edu
\IEEEcompsocthanksitem L. Pascarella is with SEART @  Software Institute, Universit\`a della Svizzera italiana, Switzerland. \protect\\
E-mail: luca.pascarella@usi.ch
\IEEEcompsocthanksitem A. Mastropaolo is with SEART @  Software Institute, Universit\`a della Svizzera italiana, Switzerland. \protect\\
E-mail: antonio.mastropaolo@usi.ch
\IEEEcompsocthanksitem E. Aghajani is with SEART @  Software Institute, Universit\`a della Svizzera italiana, Switzerland. \protect\\
E-mail: emad.aghajani@usi.ch
\IEEEcompsocthanksitem D. Poshyvanyk is with SEMERU @ William \& Mary, USA. \protect\\
E-mail: denys@cs.wm.edu
\IEEEcompsocthanksitem M. Di Penta is with University of Sannio, Italy. \protect\\
E-mail: dipenta@unisannio.it
\IEEEcompsocthanksitem G. Bavota is with SEART @  Software Institute, Universit\`a della Svizzera italiana, Switzerland. \protect\\
E-mail: gabriele.bavota@usi.ch}% <-this % stops an unwanted space
}

\markboth{Journal of \LaTeX\ Class Files,~Vol.~xx, No.~x, Month~xxxx}%
{Ciniselli \MakeLowercase{\etal}: An Empirical Study on the Usage of Transformer Models for Code Completion}
\IEEEtitleabstractindextext{%
\begin{abstract}
\justifying
Code completion aims at speeding up code writing by predicting the next code token(s) the developer is likely to write. Works in this field focused on improving the accuracy of the generated predictions, with substantial leaps forward made possible by deep learning (DL) models. However, code completion techniques are mostly evaluated in the scenario of predicting the next token to type, with few exceptions pushing the boundaries to the prediction of an entire code statement. Thus, little is known about the performance of state-of-the-art code completion approaches in more challenging scenarios in which, for example, an entire code block must be generated. We present a large-scale study exploring the capabilities of state-of-the-art Transformer-based models in supporting code completion at different granularity levels, including single tokens, one or multiple entire statements, up to entire code blocks (\eg the iterated block of a \textit{for} loop). We experimented with several variants of two recently proposed Transformer-based models, namely \roberta and the Text-To-Text Transfer Transformer (T5), for the task of code completion. The achieved results show that Transformer-based models, and in particular the T5, represent a viable solution for code completion, with perfect predictions ranging from $\sim$29\%, obtained when asking the model to guess entire blocks, up to $\sim$69\%, reached in the simpler scenario of few tokens masked from the same code statement.
\end{abstract}

}

\maketitle

\IEEEdisplaynontitleabstractindextext
\IEEEpeerreviewmaketitle

\input{intro}
\input{design}
\input{results}
\input{threats}
\input{related}
\input{conclusion}

\section*{Acknowledgment}
This project has received funding from the European Research Council (ERC) under the European Union's Horizon 2020 research and innovation programme (grant agreement No. 851720).
W\&M co-authors have been supported in part by the NSF CCF-1955853 and CCF-2007246 grants. Any opinions, findings, and conclusions expressed herein are the authors' and do not necessarily reflect those of the sponsors.

\balance

\bibliographystyle{IEEEtranS}
\bibliography{main}
\newpage

\input{bios}

\end{document}

%% file: intro.tex
% !TEX root = main.tex
%%%%%%%%%%%%%%%%%%%%%%%%%%%%%%%%%%%%%%%%
%%%%%%%%%%%%%%%%%%%%%%%%%%%%%%%%%%%%%%%%
\section{Introduction} \label{sec:intro}
%%%%%%%%%%%%%%%%%%%%%%%%%%%%%%%%%%%%%%%%
%%%%%%%%%%%%%%%%%%%%%%%%%%%%%%%%%%%%%%%%
Code completion is considered as one of the ``killer'' features of modern Integrated Development Environments (IDEs)~\cite{Bruch:fse2009,Robb2010a,kim2020code}: It can provide developers with predictions about the next code token (\eg a method call) to write given the code already written in the IDE, thus speeding up software development and preventing potential mistakes~\cite{han2009code,han2011code}.

Several works in this field have been proposed. Most of them aim at advancing the performance of code completion tools, especially in terms of prediction accuracy. Such research has allowed moving from simple alphabetically ranked lists of recommendations for completing what a developer is typing (\eg a list of possible method calls matching what has been typed by the developer) to ``intelligent'' completions considering the context surrounding the code \cite{Bruch:fse2009,Robb2010a}, the history of code changes \cite{Robb2010a}, and/or coding patterns mined from software repositories \cite{Hindle:icse2012,Nguyen:icse2012,Tu:fse2014,Asaduzzaman2014,Nguyen:msr2016,niu2017api,Hellendoorn:fse2017}. Last, but not least, Deep Learning (DL) models have been applied to code completion \cite{White2015,Karampatsis:DLareBest,kim2020code,alon2019structural,svyatkovskiy2020intellicode,Ciniselli2021}, setting new standards in terms of prediction performance. Although the performance of code completion techniques has substantially improved over time, the type of support they provide to developers has not evolved at the same pace. Indeed, besides a few works focusing on predicting multiple code tokens (\eg \cite{alon2019structural,svyatkovskiy2020intellicode}) or even recommending entire statements (\eg \cite{Asaduzzaman:icsme2017,Wen:icse2021}), most of the approaches presented in the literature have only been experimented in the specific scenario in which the next token the developer is likely to type must be predicted. This leaves the following question partially unanswered: \textit{how far can we go with DL-based token prediction (even beyond the source code line boundary)}?

We present a large-scale empirical study exploring the limits and capabilities of state-of-the-art DL models to support code completion. Besides generating the next token(s) the developer is likely to write, we apply DL models to generate entire statements and code blocks (\eg the body of an {\tt if} statement). Among the many DL models proposed in the literature, we focus on models using the Transformer architecture \cite{attention}. In particular, in our recent work published at MSR 2021 \cite{Ciniselli2021} we evaluated the performance of a \roberta model \cite{roberta} in the code completion tasks described above. \roberta is a BERT (Bidirectional Encoder Representations from Transformers) model \cite{Delvin:2019} using a pre-training task in which random words in the input sentences are masked out using a special \mask token, with the model in charge of predicting the masked words. While experimenting with \roberta for the task of code completion, we faced an important limitation that did not make it suitable for the study we wanted to perform (\ie the prediction of multiple masked tokens): In the \roberta pre-training task $n$ \mask tokens must be used to mask $n$ code tokens, thus implicitly suggesting to the model how many code tokens must be generated to autocomplete the masked statement. This would not be realistic in a real usage scenario, in which the code completion engine must \emph{guess} the tokens to generate, without the developer suggesting how many tokens must be generated. To overcome this limitation, we had to adapt the \roberta pre-training objective to be able to guess, from a single \mask token masking one or more code tokens in the given statements, which and how many code tokens must be generated \cite{Ciniselli2021}. The adaptation of the \roberta pre-training objective was inspired by the recently proposed Text-To-Text Transfer Transformer (T5) architecture \cite{raffel2019exploring}, suggesting this as a good fit for the task of code completion. 

In this work, we extend our MSR 2021 paper \cite{Ciniselli2021} by showing that the T5 substantially overcomes the performance of the \roberta model, being able to correctly predict even entire code blocks, something that we found to be not achievable with \roberta. As in \cite{Ciniselli2021}, we focus on three code prediction scenarios: (i) \emph{token-level} predictions, namely classic code completion in which the model is used to guess the last $n$ tokens in a statement the developer started writing; (ii) \emph{construct-level} predictions, in which the model is used to predict specific code constructs (\eg the condition of an {\tt if} statement) that can be particularly useful to developers while writing code; and (iii) \emph{block-level} predictions, with the masked code spanning one or more entire statements composing a code block (\eg the iterated block of a {\tt for} loop).

We compare the performance of several models. First, we use the \roberta model as presented in \cite{Ciniselli2021} as representative of BERT-like models. Second, we use T5 model for the task of code completion for the first time in this paper. The T5 has been recently shown to outperform many state-of-the-art techniques in code-related tasks \cite{Matropaolo:icse2021}. In particular,  Mastropaolo \etal \cite{Matropaolo:icse2021} showed the possibility to train a single T5 model dealing with four code-related tasks, namely bug fixing, injection of code mutant, assert statements generation, and code summarization. Third, we experiment with an \textit{n}-gram model as a baseline for DL-based models, also showing the impact on its performance of using a caching mechanism as proposed by Hellendoorn and Devanbu~\cite{Hellendoorn:fse2017}.

Both \roberta and T5 models are trained in two phases: \textit{pre-training} which allows defining a shared knowledge-base useful for a large class of sequence-to-sequence tasks (\eg guessing masked words in English sentences to learn about the language), and \textit{fine-tuning} which specializes the model on a specific downstream task (\eg learning the translation of sentences from English to German). 

Several tasks can be used in the fine-tuning, to possibly take advantage of \emph{transfer learning} (\ie the knowledge acquired on one task can be reused by the model for another task). For example, a single model trained on multiple translation tasks (\eg from English to German,  English to French, and French to German) could be more effective than three different models each trained on a specific translation task (\eg English to German).

In our work, we want to investigate the performance of the two transformer-based models by also looking at the role played on the models' performance by the pre-training task and the transfer learning across different tasks. However, since this requires the training of many different variants of the experimented models, we adopt the following strategy. First, we compare \roberta and T5 by training three different models for the three code completion scenarios (\ie token-level, construct-level, and block-level) we experiment with. This implies creating three different \roberta and T5 models (six models overall). Then, we take the best performing one (T5) and we show that using pre-training increases its performance, even though the impact is limited. Finally, we show that fine-tuning a single T5 model to support all three prediction tasks boosts performance confirming transfer learning across the three very similar tasks (\ie knowledge acquired in one task can be used to perform another task). 

The achieved results show that, for a typical code completion task (\ie \emph{token-level}), T5 correctly guesses all masked tokens in 66\% to 69\% of cases (depending on the used dataset), while \roberta achieving 39\% to 52\% and the $n$-gram model 42\% to 44\%. In the most challenging prediction scenario, in which we mask entire blocks, \roberta and the $n$-gram model show their limitations, being able to only correctly reconstruct the masked block in less than 12\% of the cases, while the T5 achieves 30\% of correct predictions.

It is worth noting that the goal of our study is not to show that the T5 model is the best option for neural-based code completion. Our work focuses on empirically exploring the capabilities of learning-based code completion techniques, and T5, \roberta, and the $n$-gram model have been chosen as representatives of the state-of-the-art.

In summary, as compared to our MSR 2021 paper \cite{Ciniselli2021}, the contributions of this work are as the following: (i) we perform a comprehensive empirical study with an additional state-of-the-art approach, namely the T5 model, showing its very promising performance for the code completion task; (ii) differently from \cite{Ciniselli2021} in which three different \roberta models have been fine-tuned on the three code completion scenarios (\ie token-level, construct-level, and block-level) without pre-training and without testing the impact of transfer learning, we pre-train and fine-tune several versions of the best performing model (\ie the T5), to investigate these aspects; (iii) for the best performing model, we also explore the possibility of exploiting the confidence of the predictions as a measure of the prediction quality, showing the reliability of such an indicator.

The source code and data used in our work are publicly available in a comprehensive replication package \cite{replication}.

%% file: design.tex
% !TEX root = main.tex
%%%%%%%%%%%%%%%%%%%%%%%%%%%%%%%%%%%%%%%%
%%%%%%%%%%%%%%%%%%%%%%%%%%%%%%%%%%%%%%%%
\section{Research Questions and Context} \label{sec:design}
%%%%%%%%%%%%%%%%%%%%%%%%%%%%%%%%%%%%%%%%
%%%%%%%%%%%%%%%%%%%%%%%%%%%%%%%%%%%%%%%%
The study \emph{goal} is to assess the effectiveness of Transformer-based DL models in predicting masked code tokens at different granularity levels. We address the following research questions (RQs):

{\bf RQ$_1$:} \emph{To what extent are transformer models a viable approach to learn how to autocomplete code?} This RQ investigates the extent to which T5 and \roberta can be used for predicting missing code tokens. We assess the quality of the generated predictions from both a quantitative (\ie BLEU score \cite{dreyer:bleu}, Levenshtein distance \cite{levenshtein1966}) and a qualitative (\ie perfect predictions, potential usefulness of wrong predictions) point of view. RQ$_1$ is further detailed in the following two sub-RQs:

{\bf RQ$_{1.1}$:} \emph{To what extent does the number of masked tokens impact the prediction quality?} We train and test the approaches we experiment with on datasets in which masked code tokens span from few contiguous tokens in a given statement to multiple missing statements composing a code block. RQ$_{1.1}$ explores the limits of Transformer models when considering simple and more challenging code completion scenarios.

{\bf RQ$_{1.2}$:} \emph{To what extent are the performance of the models influenced by the specificity of the dataset employed for training and testing it?} While it is reasonable to expect that larger training datasets tend to help deep learning models, we are interested in answering RQ$_{1.2}$ from a different perspective. To address this RQ, we compare the autocompletion performance on two different datasets: a first, more general one, composed of Java methods; and a second, more specific one, composed of methods from Android apps. While the programming language is the same, \and the granularity of the two datasets is the same (\ie method-level granularity), methods in the second dataset make heavy use of Android APIs, and the same APIs are likely to be used for similar purposes, \eg app features dealing with GPS positioning share common API usages. We expect this to create ``regularities'' in the Android dataset to help model learning. 

{\bf RQ$_{2}$:} \emph{What is the role of pre-training and transfer learning in the performance of Transformer-based models?} As explained in \secref{sec:intro}, both \roberta and T5 can be pre-trained and then fine-tuned on several tasks. RQ$_{2}$ investigates the boost in performance (if any) brought by (i) pre-training of the models, and (ii) fine-tuning a single model on several tasks to take advantage of transfer learning. Such an additional analysis has been performed only for the best-performing model (\ie the T5).

{\bf RQ$_3$:} \emph{How do transformer models compare to a state-of-the-art $n$-gram model?} An alternative to DL models is represented by statistical language models based on \textit{n}-grams. In this research question, we compare the DL models to (i) a classical $n$-gram model and, (ii) in a smaller study, to the state-of-the-art \textit{n}-gram cached model \cite{Hellendoorn:fse2017}. 

\subsection{Context Selection: Datasets}
\label{sec:datasets}
Our study involves two datasets. The first one comes from our MSR'21 paper \cite{Ciniselli2021} and is used to fine-tune the \roberta and T5 models and to train the $n$-gram model. We refer to this dataset as \emph{fine-tuning dataset} and it includes both a Java and an Android dataset to allow answering RQ$_{1.2}$. The fine-tuning dataset has been built starting from the CodeSearchNet dataset \cite{Java:CodeSearchNet}, which features Java methods mined from open source projects. The second dataset has been built specifically to answer RQ$_{2}$, \ie to have a different dataset that can be used to pre-train the best performing model among \roberta and T5 (\ie \emph{pre-training dataset}). The following section describes how the datasets have been built. 

\subsubsection{Fine-tuning dataset}
\label{sub:fine-tuning}
To create the \emph{Java dataset}, we started from the CodeSearchNet Java Dataset provided by Husain \etal \cite{Java:CodeSearchNet}.  We decided to start from CodeSearchNet rather than from other datasets proposed in the literature (see \eg \cite{50k,maven-data}) since CodeSearchNet has been already subject to cleaning steps making it suitable for applications of machine learning on code. Also, CodeSearchNet is already organized at method-level granularity (\ie one instance is a method), while other datasets, such as the 50k~\cite{50k}, collect whole repositories. In particular, CodeSearchNet contains over 1.5M Java methods collected from open-source, non-fork, GitHub repositories. For details on how the dataset has been built, see the report by Husain \etal \cite{Java:CodeSearchNet}. For our work, the most important criteria used in the dataset construction are: (i) excluding methods of fewer than three lines; (ii) removing near-duplicate methods using the deduplication algorithm from CodeSearchNet; this is done to not inflate the performance of the models as a result of overlapping instances between training and test sets \cite{Deduplication} and (iii) removing methods with the name containing the ``test'' substring in an attempt to remove test methods; methods named ``toString'' are removed as well. The latter are often automatically generated by the IDEs with a very similar structure (\eg mostly concatenating class attributes). Thus, they rarely represent a challenging code completion scenario and can result in inflating the prediction accuracy. 

To build the \emph{Android dataset} we adopted a similar procedure. We cloned the set of 8,431 open-source Android apps from GitHub available in the AndroidTimeMachine dataset \cite{Geiger:2018}. Then, we extracted from each project's latest snapshot the list of methods. This resulted in a total of $\sim$2.2M methods. Then, we applied the same filtering heuristics defined for the Java dataset, ending up with 654,224 methods. Since one of the goals of our study is also to compare the performance of the models when applied on a more generic (Java) and a more specific (Android) dataset, we randomly selected (using the \texttt{random} Python function) 654,224 methods from the Java dataset, to match the size of the Android dataset.

In our MSR paper \cite{Ciniselli2021}, we also experimented with code abstraction as used in the previous studies \cite{Tufano:icse2019,Tufano:tosem2019} to avoid the open vocabulary problem. However, new DL-based models do not suffer from this limitation anymore thanks to the usage of tokenizers exploiting techniques such as Byte Pair Encoding (BPE) \cite{bpe}. For this reason, while in \cite{Ciniselli2021} we built two versions of the fine-tuning dataset (with and without abstraction), in this work we only focus on the datasets using raw source code since this is the real scenario in which code completion techniques are used. Such clarification is needed since, when building the \emph{fine-tuning dataset}, methods for which parsing errors occurred during the abstraction process were excluded \cite{Ciniselli2021}, leaving the Java dataset with 634,799 methods, and the Android one with 532,096.

Then, the three versions of each dataset (Java and Android) summarized in \tabref{tab:dataset} were created using the following masking processes (note that \tabref{tab:dataset} reports the number of instances in each dataset after the filtering steps described below):

\textbf{Token masking.} For each code line $l$ in each method having more than one token we mask its last $x$ tokens, where $x$ is a random number between $1$ $\dots$ $n-1$, where $n$ is the number of tokens composing $l$. Given a method $m$ having $k$ lines with more than one token, we generate $k$ versions of $m$, each of them having one line with the last $x$ tokens masked and the remaining $k-1$ reported without any change (\ie no masked tokens, just the original raw source code). We set the maximum number of masked tokens to 10 (\ie if $x > 10$ then $x=10$). This scenario simulates the classic code completion task in which a developer is writing a statement and the code completion tool is in charge of autocompleting it.

\textbf{Construct masking.} We selected a number of code constructs for which it could be particularly useful to be supported with automated code completion. Given a method $m$, we use the scrML \cite{SrcML} toolkit to identify all $m$'s tokens used to: (i) define the complete condition of an {\tt if} statement or of a {\tt while}/{\tt for} loop (\eg in a statement having {\tt for(int i=0; i<data.size(); i++)} we identify all tokens between parenthesis as those used to define the {\tt for} loop); (ii) define the parameters in a method call (\eg in {\tt copyFile(source, target)} the tokens ``{\tt source}'', ``{\tt ,}'', and ``{\tt target}'' are identified); and (iii) define the exception caught in a {\tt catch} statement (\eg in {\tt catch(IOException io)} we identify {\tt IOException io} as the involved tokens). For $m$ this results in a set $S$=\{$T_1$, $T_2$, $\dots$, $T_n$\}, where $T_i$ represents a set of relevant tokens for one of the previously mentioned constructs (\eg $T_i$ is the set of tokens used to define the {\tt for } loop condition). 

Given $m$, we generate $|S|$ versions of it, each one having one of the subject constructs masked. Also, in this case we set the maximum number of masked tokens to 10. This means that if a construct requires more than 10 tokens to be masked (this happened for 9.38\% of the constructs in our dataset), it is not masked in our dataset.

The code completion tasks simulated by the construct masking resemble cases in which the developer uses the technique to get recommendations about non-trivial code tokens, representing decision points in the program flow (\eg condition of \texttt{if} statement) or error-handling cases (\eg exceptions to catch).

\textbf{Block masking.} We use srcML to identify in each method $m$ its code blocks. We define a code block as the code enclosed between two curly brackets. For example, a block may be, besides the method body itself, the code executed in a {\tt for}/{\tt while} loop, when an {\tt if}/{\tt else}/{\tt else if} condition is satisfied, \etc Then, given $k$ the number of blocks identified in $m$, we create $k$ versions of $m$ each one having a specific code block masked. We set the maximum size of the masked block to two complete statements. This means that if a block is composed of more than two statements (which happened for 49.29\% of the blocks in our dataset), it is not masked. This is the most challenging code completion scenario in which we test the experimented techniques. If successful in this task, code completion techniques could substantially speed up code implementation activities.

\input{tables/design-tables}

In summary, there are six fine-tuning datasets: For each of the two domains (Java or Android), there are three different masking levels (token, construct, block). These masking levels have been pick to simulate code completion tasks having different complexity (with \emph{token masking} expected to be the simplest and \emph{block-masking} the most complex). 

Starting from the six datasets, we created the training, evaluation, and test sets in \tabref{tab:dataset}. As a first step, we filtered out specific instances from our datasets. First, when using generative deep learning models, the variability in length of the sentences (in our case, methods) provided as input can affect the training and performance of the model, even when techniques such as padding are employed. For this reason, we analyzed the distribution of methods length in our dataset, finding that two-thirds of them are composed of at most 100 tokens. For this reason, as done by Tufano \etal~\cite{Tufano:tosem2019}, we excluded from our datasets all the methods having more than 100 tokens. Second, \roberta cannot efficiently handle cases in which the \emph{masked} tokens are more than the \emph{non-masked} tokens. This often happens, for example, when masking the entire method body in the block-level masking approach. Thus, those instances are excluded as well. 

After the filtering steps, we split each of the six datasets into training (80\%), evaluation (10\%), and test (10\%) sets. While the methods in the dataset are randomly ordered, the splitting we performed was not random to avoid biasing the learning. To explain this point, let us consider the case of the \emph{block masking} dataset. Given a method $m$ having $k$ blocks in it, we add in the dataset $k$ versions of $m$, each having one and only one block masked. Suppose that $m$ contains two blocks $b_1$ and $b_2$, thus leading to two versions of $m$: One in which $b_1$ is masked ($m_{b_1}$) and $b_2$ is not and \emph{vice versa} ($m_{b_2}$). With a random splitting, it could happen that $m_{b_1}$ is assigned to the training set and $m_{b_2}$ to the test set. However, in $m_{b_1}$ the $b_2$ is not masked. Thus, when the model has to guess the tokens masked in $m_{b_2}$ it would have the solution in the training set, resulting in boosted prediction performance. For this reason, we randomly select 80\% of the methods in each dataset and assign all of their masked versions to the training set. Then, we proceed in the same way with evaluation and test sets. 

Using this procedure, we obtained the datasets in \tabref{tab:dataset}. Important to note is that, given the original size of the datasets using token-level and construct-level masking, we decided to cap the training set to 750k instances (no changes were done in the evaluation and test sets). This was necessary given the computationally expensive process of training several DL models (as it will be clear later, our study required the training of 19 different DL-based models). Also, the size of the evaluation and test sets is slightly different since, as explained before, we split the dataset based on the methods (not on their masked versions) and each method can result in a different number of its generated masked versions.

\subsubsection{Pre-training dataset}
To build the pre-training dataset, we used the GitHub Search platform \cite{Dabic:msr2021data} to identify all Java repositories having at least 100 commits, 10 contributors, and 10 stars. These filtering criteria only aim at reducing the likelihood of selecting toy and personal projects for the building of this dataset. We sorted the projects by their number of stars, cloning the top 6,000 and extracting from each of them the methods in the latest snapshot tagged as a release, to only rely on methods likely to be syntactically correct. Repositories for which no snapshot was tagged as a release were excluded, leaving 3,175 repositories. Finally, since we wanted to avoid extremely large projects to influence the dataset too much (\ie to contribute too many methods to the dataset), we cap the maximum number of methods to extract from each repository to 1,500. This was also due to limit the number of the pre-training instances to a manageable size according to our available hardware resources. In addition to the filters used while building the fine-tuning dataset (see \secref{sub:fine-tuning}), we also removed test methods identified as all those using the \textit{@test} annotation or containing the word ``test'' in the method name after camel case splitting (\ie we do not exclude \textit{upda\underline{teSt}atus}). Also, since the goal of the pre-training dataset is to provide instances in which random tokens are masked to make the model ``familiar'' with a specific context (\ie the Java language in our case), we excluded very short methods ($<$ 15 tokens) not having enough elements to mask and, for the same reasons explained for the fine-tuning dataset, long methods (in this case, $>$ 200 tokens). 

We then removed all the exact duplicates within the pre-training dataset, keeping in the dataset only the first occurrence of each duplicate. After having removed the duplicates, the dataset contained 1,874,338 different methods. Finally, we ensured that the pre-training dataset does not contain any methods belonging to the fine-tuning dataset (neither in the training, evaluation, or test sets). We found a total of 23,977 duplicates between the pre-training and the fine-tuning datasets, leading to a final number of 1,850,361 instances in the \emph{pre-training} dataset.

\subsection{Context Selection: Techniques}
In this section we overview three experimented techniques, \ie \roberta \cite{roberta}, T5 \cite{raffel2019exploring}, and \ngram \cite{Hellendoorn:fse2017}. We refer to the original papers presenting them for additional details. 

\subsubsection{\roberta}

The first Transformer-based model leverages the off-the-shelf \roberta model, which is an Encoder-Transformer architecture. Details about the \roberta model are provided in a report by Liu \etal \cite{roberta}, while here, we mainly focus on explaining why it represents a suitable choice for code completion. BERT-based models, such as \roberta, use a special pre-training where random words in the input sentence are masked out with a special \mask token. This pre-training task is very well-suited to simulate a code completion task, in which the input is an incomplete code snippet the developer is writing and the masked tokens represent the code needed to autocomplete the snippet. However, one limitation of such a pre-training is that when attempting to predict multiple tokens, \eg an entire masked \textit{if} condition, it requires the number of tokens to generate to be known, due to the fixed sequence length of Transformers \cite{attention}. To overcome this issue, we modify such an objective by masking spans of tokens using a single \mask token. 

As previously explained, BERT models (such as \roberta) can be pre-trained and fine-tuned on several tasks \cite{Delvin:2019}. The result will be a single model able to support different tasks and, possibly, taking advantage of what it learned for a specific task to also improve its performance in a different task. In our study, we start by comparing the \roberta and the T5 models in a scenario in which no pre-training is performed and a single model is built for each of the three code completion tasks previously described (\ie token, construct, and block masking) by using the \emph{fine-tuning dataset}. Then, for the best performing model among the two (\ie T5), we also experiment with pre-training and multi-task fine-tuning.  We trained six \roberta models, one for each dataset in \tabref{tab:dataset}.

As for the implementation of the \roberta model, we used the one provided in the Python \emph{transformers} library~\cite{Wolf2019HuggingFacesTS}.
We also train a tokenizer for each model to overcome the \emph{out-of-vocabulary problem}. The \emph{out-of-vocabulary problem} happens when a machine learning model deals with terms that were not part of the training set but appear in the test set. We trained a Byte Pair Encoding (BPE) \cite{bpe} model using the HuggingFace's \emph{tokenizers} Python library~\cite{tokenizers}. BPE uses bytes as vocabulary, allowing it to tokenize every text without requiring the unknown token often used in applications of DL to NLP, thus overcoming the \emph{out-of-vocabulary problem}. When used on source code\cite{karampatsis2020big}, BPE has been shown to address the \emph{out-of-vocabulary problem}.

\subsubsection{T5}

Raffel \etal~\cite{raffel2019exploring} presented the T5 model that leverages multi-task learning to implement \emph{transfer learning} in the NLP domain. The T5 has been presented in five pre-defined variants \cite{raffel2019exploring}: small, base, large, 3 Billion, and 11 Billion that differ in complexity, size, and, as a consequence, training time. T5$_{small}$, the smaller variant, has 60 million parameters while T5$_{11B}$, the largest, has 11 billion parameters. Despite Raffel \etal \cite{raffel2019exploring} report that highlights the largest model offers the best accuracy, its training time is sometimes too high to justify its use. Given our computational resources, we opted for the T5$_{small}$ model; therefore, we expect that our results represent a lower bound for the performance of a T5-based model. 

T5 offers two advantages as compared to other DL models: (i) it is usually more efficient than RNNs since it allows to compute the output layers in parallel, and (ii) it can detect hidden and long-ranged dependencies among tokens, without assuming that nearest tokens are more related than distant ones. The latter is particularly relevant in code-related tasks. For example, a local variable could be declared at the beginning of a method (first statement), used in the body inside an \texttt{if} statement, and finally returned in the last method's statement. Capturing the dependency existing between these three statements, that might even be quite far from each other (\eg variable declaration and return statement), can help in better modeling the source code with a consequent boost of performance for supporting code-related tasks.

For additional details about the T5 architecture, we refer the reader to the original work presenting this model \cite{raffel2019exploring}.

\subsubsection{\ngram}
As a baseline for comparison, we used the widely studied statistical language models based on \ngram. An \ngram model can predict a single token following the $n-1$ tokens preceding it. Even though the $n$-gram model is meant to predict a single token given the $n-1$ preceding tokens, we designed a fair comparison for the scenario in which we mask more than one token. In particular, we use the $n$-gram model in the following way: Let us assume that we are predicting, using a $3$-gram model, how to complete a statement having five tokens \texttt{T}, of which the last two are masked (\texttt{M}): $<$\texttt{T$_1$}, \texttt{T$_2$}, \texttt{T$_3$}, \texttt{M$_4$}, \texttt{M$_5$}$>$, with \texttt{M$_4$} and \texttt{M$_5$} masking \texttt{T$_4$} and \texttt{T$_5$}, respectively. We provide as input to the model \texttt{T$_2$} and \texttt{T$_3$} to predict \texttt{M$_4$}, obtaining the model prediction \texttt{P$_4$}. Then, we use \texttt{T$_3$} and \texttt{T$_4$} to predict \texttt{M$_5$}, thus obtaining the predicted sentence $<$\texttt{T$_1$}, \texttt{T$_2$}, \texttt{T$_3$}, \texttt{P$_4$}, \texttt{P$_5$}$>$. Basically, all predictions are joined to predict multiple contiguous tokens.

The \ngram models are trained on the same training sets used for the fine-tuning of the DL models without, however, masked tokens. We experiment with both the standard $n$-gram model (\ie the one discussed above) as well as, in a smaller study, with the \ngram cached model proposed by Hellendoorn and Devanbu \cite{Hellendoorn:fse2017}.

\section{Data Collection and Analysis}
In this section we detail the data collection and analysis procedure adopted to answer the research questions described in \secref{sec:design}.

\subsection{Training of Models}
We detail the process used for the training and hyperparameters tuning of the two deep learning models that we experimented with.

\input{tables/hyperparameters}

\subsubsection{\roberta}
We performed hyperparameter tuning using the Weights \& Biases's \cite{wandb} Python library on a Linux server with an Nvidia RTX Titan GPU. \tabref{tab:hyperRoberta} reports the hyperparameters we tuned, the range of values we tested for them, and the value in the best configuration we found. Besides those parameters, we used an attention dropout probability of 0.1, and a hidden layer dropout probability of 0.3. For the tokenizer, the vocabulary size was set to 50k. The hyperparameter search was performed using the training and the evaluation sets of the Android dataset with token masking. We picked as the best configuration the one that, when applied to the evaluation set, was able to obtain the highest number of ``perfect predictions''. We define as ``perfect'' a prediction that exactly matches the code written by the developers. Thus, the model correctly guesses \emph{all} masked tokens. If one of the masked tokens is different we do not consider the prediction as ``perfect''. While, in principle, a different hyperparameter tuning would be necessary for each dataset, such a process is extremely expensive, and preliminary investigations we performed on a subset of the other datasets showed minor differences in the achieved best configuration.

The training was performed across servers using their GPUs. The first was equipped with an Nvidia Tesla V100S, the second with an Nvidia RTX Titan, and the third with 3 Nvidia GTX 1080Ti. The training time strongly depends on the size of the dataset and the used server but ranged between 28 and 114 hours per model. Note that, once trained, each model can be used to perform predictions in the split of a second (on average, 0.12 seconds on a laptop CPU), thus making them a viable solution for ``real-time'' code completion.

We train each model for a maximum of 50 epochs. However, we adopted the following stopping condition. At the end of each training epoch, we executed the model on the evaluation set and we compute the number of perfect predictions. If we observe that, during the training, the performance of the model is worsening in terms of perfect predictions on the evaluation set (\ie the model is likely overfitting to the training set), we stop the training. In particular, given a model trained for $n^{th}$ epoch, we stop the training if the number of perfect predictions on the evaluation set is lower than the number of perfect predictions achieved after the $n-4$ epoch. This ensures that the models can have some fluctuations in performance for up to three epochs. Then, if it is still not improving, we stop its training and take the best model (in terms of perfect predictions on the evaluation test) obtained up to that moment. None of the models were trained for the whole 50 epochs.

\begin{table}[h]
	\centering
	\caption{Hyperparameters Tuned for the T5 Models.}
	\begin{tabular}{ll}
		\toprule
		\textbf{Learning Rate Type} & \textbf{Parameters}                             \\
		\midrule
		Constant (C-LR)             & $\mathit{LR} = 0.001$                           \\
		Slanted Triangular (ST-LR)  
		                            & $\mathit{LR}_{\mathit{starting}} = 0.001$       \\
		                            & $\mathit{LR_{\mathit{max}}} = 0.01$             \\
		                            & $\mathit{Ratio} = 32$                           \\
		                            & $\mathit{Cut} = 0.1$                            \\
		Inverse Square Root (ISQ-LR) 
		                            & $\mathit{LR}_{\mathit{starting}} = 0.01$        \\
		                            & $\mathit{Warmup} = 10,000$                      \\
		Polynomial Decay (PD-LR)    
		                            & $\mathit{LR}_{\mathit{starting}} = 0.01$        \\
		                            & $\mathit{LR}_{\mathit{end}} = 1\mathrm{e}{-06}$ \\
		                            & $\mathit{Power} = 0.5$                          \\
		\bottomrule
	\end{tabular}
	\label{tab:hyperT5}
\end{table}

\subsubsection{T5} \label{sec:design:training_t5}
We rely on the same configurations used by Mastropaolo \etal \cite{Matropaolo:icse2021}. In particular, concerning the pre-training, we do not tune the hyperparameters of the T5 model because the pre-training step is task-agnostic, and this would provide limited benefits. Instead, we experiment with four different learning rate schedules for the fine-tuning phase, using the configurations reported in \tabref{tab:hyperT5}, and identify the best-performing configuration in terms of perfect predictions on the evaluation sets. Each of the four experimented configurations has been trained for 100k steps ($\sim$7 epochs) before assessing its performance on the evaluation sets. Across all six evaluation datasets (\tabref{tab:dataset}), the best performing configuration was the one using the Slanted Triangular learning rate, confirming the findings in \cite{Matropaolo:icse2021}. Also, all T5 models we built use a \emph{SentencePiece} \cite{DBLP:journals/corr/abs-1808-06226} tokenizer trained on the pre-training dataset and are composed of 32k word pieces \cite{Matropaolo:icse2021}.

The best configuration we identified has been used to train six different T5 models (\ie one for each dataset in \tabref{tab:dataset}) and assess their performance on the corresponding test set. These results can be used to compare directly the T5 and the \roberta model when fine-tuned without pre-training and in a single-task setting (\ie no transfer learning). Since we found the T5 to perform better than \roberta, we also use this model to answer RQ$_{2}$. Thus, in addition to these six models, we also built additional seven models: six of them leverage pre-training plus single-task fine-tuning. In other words, they are the equivalent of the first six models we built, with the addition of a pre-training phase. 

For pre-training the T5 model, we randomly mask $15\%$ of the tokens in each instance (method) of the \emph{pre-training dataset}. The pre-training has been performed for 200k steps ($\sim$28 epochs), since we did not observe any improvement going further. We used a 2x2 TPU topology (8 cores) from Google Colab to train the model with a batch size of 256, with a sequence length (for both inputs and targets) of 256 tokens. As a learning rate, we use the \emph{Inverse Square Root} with the canonical configuration \cite{raffel2019exploring}.
The training requires around 26 seconds for 100 steps.

Finally, we created a T5 model exploiting both pre-training and multi-task fine-tuning (\ie a single model was first pre-trained, and then fine-tuned on all six datasets in \tabref{tab:dataset}). This was done to check the impact of transfer learning on the model performance. Overall, we trained 13 T5 models: six with no pre-training and single-task fine-tuning, six with pre-training and single-task fine-tuning, and one with pre-training and multi-task fine-tuning.

\begin{table*}[h!]
\centering
\caption{Summary of the evaluation metrics used in our study}
\label{tab:metrics}
\begin{tabular}{ll}
\hline
\textbf{Metric} & \textbf{Purpose}\\ \hline
BLEU score & Overall prediction quality for different prediction lengths\\
Levenshtein distance & Proxy of the effort needed to adapt the prediction\\
\% of perfect predictions & To what extent is the approach able to generate predictions that need human intervention\\
\hline
\end{tabular}
\end{table*}

\subsection{Analysis of Results}
\label{sec:analysis}
To answer RQ$_1$ we compute the metrics summarized in \tabref{tab:metrics} by running each trained model on the test sets in \tabref{tab:dataset}.

The first metric, \emph{Bilingual Evaluation Understudy (BLEU)-n score}, assesses the quality of automatically translated text \cite{dreyer:bleu}. The BLEU score computes the weighted percentage (\ie considering the number of occurrences) of words appearing in translated text and the reference text. We use four variants of BLEU, namely BLEU-1, BLEU-2, BLEU-3, and BLEU-4. A BLEU-n variant computes the BLEU score by considering the n-grams in the generated text. Most of the previous work in the SE literature adopts the BLEU-4 score  \cite{Gu:2016,Jiang:ASE'17,Watson:icse2020}. However, such a variant cannot be computed when the target prediction (in our case, the number of masked tokens) is lower than four. For this reason, we compute four different versions from BLEU-1 to BLEU-4. BLEU-1 can be computed for all predictions, while BLEU-n with n$>$1 only for predictions having a length (\ie number of tokens) higher or equal than $n$. The BLEU score ranges between 0\% and 100\%, with 100\% indicating, in our case, that the code generated for the masked tokens is identical to the reference one.

\emph{The Levenshtein distance \cite{levenshtein1966}}. To provide a proxy measure of the effort needed by developers to convert a prediction generated by the model into the reference (correct) code, we compute the Levenshtein distance at token-level: This can be defined as the minimum number of token edits (insertions, deletions or substitutions) needed to transform the predicted code into the reference one. Since such a measure is not normalized, it is difficult to interpret it in our context. Indeed, saying that five tokens must be changed to obtain the reference code tells little without knowing the number of tokens in the reference code. For this reason, we normalize such a value by dividing it by the number of tokens in the longest sequence among the predicted and the reference code.

\emph{The percentage of perfect predictions} tells us about the cases in which the experimented model can recommend the very same sequence of tokens which were masked in the target code.

We statistically compare the results achieved by \roberta and T5 using different statistical analyses. We assume a significance level of 95\%. As explained below, we use both tests on proportions and non-parametric tests for numerical variables; parametric tests cannot be used because all our results in terms of BLEU score or Levenshtein distance deviate from normality, according to the Anderson-Darling test \cite{adtest} (\emph{p}-values$<$0.001). Whenever an analysis requires running multiple test instances, we adjust \emph{p}-values using the Benjamini-Hochberg procedure \cite{bh}.

To (pairwise) compare the perfect predictions of \roberta and T5, we use the McNemar's test \cite{mcnemar}, which is a proportion test suitable to pairwise compare dichotomous results of two different treatments. To compute the test results, we create a confusion matrix counting the number of cases in which (i)  both T5 and \roberta provide a perfect prediction, (ii) only T5 provides a perfect prediction, (iii) only \roberta provides a perfect prediction, and (iv) neither T5 nor \roberta provides a perfect prediction. Finally, we complement the McNemar's test with the Odds Ratio (OR) effect size.

The comparison between different datasets, aimed at addressing RQ$_{1.2}$, is performed, again, through a proportion test, but this time, being the analysis unpaired (\ie we are comparing results over two different datasets), we use Fischer's exact test (and related OR) on a matrix containing, for different approaches and for different masking levels, the number of correct and incorrect predictions achieved on Java and Android.

To compare results of T5 and \roberta in terms of BLEU-n score and Levenshtein distance, we use the Wilcoxon signed-rank test \cite{wilcoxon} and the paired Cliff's delta \cite{Cliff:2005} effect size.
Similarly, the comparison between datasets in terms of BLUE-n score and Levenshtein distance, being unpaired, is performed using the Wilcoxon rank-sum test  \cite{wilcoxon} and the unpaired Cliff's delta effect size.

For the T5, we also statistically compare the performance achieved (i) with/without pre-training, and (ii) with/without transfer learning. Also in this case, McNemar's test is used to compare perfect predictions.

Finally, we take the best performing model (\ie T5 with pre-training and multi-task fine-tuning) and we check whether the \emph{confidence} of the predictions can be used as a reliable proxy for the ``quality'' of the predictions. If this is the case, this means that in a recommender system built around the trained model, the developer could decide to receive recommendations only when their confidence is higher than a specific threshold. T5 returns a score for each prediction, ranging from minus infinity to 0. This score is the log-likelihood ($ln$) of the prediction. Thus, if it is 0, it means that the likelihood of the prediction is 1 (\ie the maximum confidence, since $ln(1) = 0$), while when it goes towards minus infinity, the confidence tends to 0.

We split the predictions performed by the model into ten intervals, based on their confidence $c$ going from 0.0 to 1.0 at steps of 0.1 (\ie first interval includes all predictions having a confidence $c$ with 0 $\leq$ c $<$ 0.1, last interval has 0.9 $\leq$ c). Then, we report for each interval the percentage of perfect predictions.

To corroborate our results with a statistical analysis, we report the OR obtained by building a logistic regression model relating the confidence (independent variable) with the extent to which the prediction achieved a perfect prediction (dependent variable). Given the independent variable estimate $\beta_i$ in the logistic regression model, the OR is given by $e^{\beta_i}$, and it indicates the odds increase corresponding to a unit increase of the independent variable. We also determine the extent to which the confidence reported by the model correlates with the number of masked tokens. To this extent, we use the Kendall's correlation \cite{kendall1938measure}, which does not suffer from the presence of ties (occurring in our dataset) as other non-parametric correlations.

To address RQ$_3$, for all the datasets, we compare the performance of the DL-based models with that of an \textit{n}-gram model. In particular, we perform a first large-scale comparison using a standard \textit{n}-gram language model and, on a smaller dataset, we also compare the experimented techniques with the state-of-the-art cached \textit{n}-gram model~\cite{Hellendoorn:fse2017} using the implementation made available by the authors \cite{ngram}. We detail later why the cached \textit{n}-gram model was too expensive to run on the entire dataset. 

We tried to design a fair comparison, although the \textit{n}-gram model is designed to predict a single token given the $n$ tokens preceding it. Thus, in a scenario in which we mask more than one token, we use the $n$-gram model in the following way: We run it to predict each masked token in isolation. Then, we join all predictions to generate the final string (\ie the set of previously masked tokens). The $n$-gram models are trained on the same training sets used for the fine-tuning of the DL-based models without, however, masked tokens. We compare the three approaches in terms of perfect predictions generated on the test sets. A statistical comparison is performed using the McNemar's test \cite{mcnemar} and ORs. 

%% file: tables/design-tables.tex
% !TEX root = ../main.tex

% Design tables. 

\begin{table}[ht]
	\centering
	\caption{Study datasets. One instance corresponds to a method with masked token(s).}
	\scriptsize
	\label{tab:dataset}
	\begin{tabular}{lllrr}
	\toprule
	\multirow{2}{*}{\bf Domain} & {\bf Masking} & \multirow{2}{*}{\bf Dataset} & \multirow{2}{*}{\bf \#Instances} & \multirow{2}{*}{\bf \#Tokens}\\ 
	& {\bf Level} & \\\midrule
	                      & \multirow{3}{*}{Token} & Training & 750k & 46.9M\\
	                      &                                & Evaluation & 215k & 13.4M\\
	                      &                                & Test & 219k & 13.6M\\
	                      \addlinespace[0.08cm]
	                      & \multirow{3}{*}{Construct} & Training & 750k & 48.2M\\
	               Java   &                                    & Evaluation & 104k & 6.7M\\
	                      &                                    & Test & 106k & 6.7M\\
	                      \addlinespace[0.08cm]
	                      & \multirow{3}{*}{Block} & Training & 298k &19.1M\\
	                      &                                & Evaluation & 39k & 2.5M\\
	                      &                                & Test & 40k & 2.5M\\
	                      \addlinespace[0.08cm]\hline\addlinespace[0.08cm]
	                      
	                         & \multirow{3}{*}{Token} & Training & 750k&47.4M\\
	                         &                                & Evaluation & 198k &12.5M\\
	                         &                                & Test &201k &12.6M\\
	                         \addlinespace[0.08cm]
	                         & \multirow{3}{*}{Construct} & Training &750k & 48.9M\\
	             Android     &                                    & Evaluation & 99k & 6.4M\\
	                         &                                    & Test & 101k &6.5M\\
	                         \addlinespace[0.08cm]
	                         & \multirow{3}{*}{Block} & Training & 205k &13.4M\\
	                         &                                & Evaluation & 27k &1.7M\\
	                         &                                & Test & 27k & 1.8M\\
\bottomrule
\end{tabular}
\vspace{-0.2cm} 
\end{table}

%% file: tables/hyperparameters.tex
% !TEX root = ../main.tex

% Design tables. 

\begin{table}[ht]
	\centering
	\caption{Hyperparameters Tuned for the \roberta Models.}
	\scriptsize
	\label{tab:hyperRoberta}
	\begin{tabular}{lll}
	\toprule
	{\bf Hyperparameter} & {\bf Experimented Values} & {\bf Best}\\\midrule
	Learning rate & \{$5e^{-5}$, $3e^{-5}$, $2e^{-5}$\} & $5e^{-5}$\\
	Batch size & \{16, 32, 64\} & 64\\
	\# Hidden Layers & \{6, 12, 16\} & 12\\
	\# Attention Heads & \{6, 12, 16\} & 16\\
	Hidden Layer Size & \{256, 512, 768, 1024\} & 768\\
	Intermediete Size & \{3072, 4096\} & 4,096\\
\bottomrule
\end{tabular} 
\end{table}

%% file: results.tex
% !TEX root = main.tex
%%%%%%%%%%%%%%%%%%%%%%%%%%%%%%%%%%%%%%%%
%%%%%%%%%%%%%%%%%%%%%%%%%%%%%%%%%%%%%%%%
\section{Results Discussion} \label{sec:results}
%%%%%%%%%%%%%%%%%%%%%%%%%%%%%%%%%%%%%%%%
%%%%%%%%%%%%%%%%%%%%%%%%%%%%%%%%%%%%%%%%
We start by contrasting the performances of T5 and \roberta (\secref{sec:rq1}). Then, we show how the $n$-gram model compares with the DL-based ones (\secref{sec:rq2}). Finally, \secref{sec:rq1_qualitative} presents qualitative examples of correct predictions made by the models and discusses the semantic equivalence of non-perfect predictions.

Note that, upon interpreting the achieved results, and especially those concerning the perfect (correct) predictions, a direct comparison with the results achieved in previous works on code completion is not possible. This is because most of the studies in the literature experiment with code completion models when predicting a single next token the developer is likely to write. As we will show, in such a specific scenario the models we experiment with can achieve extremely high accuracy ($>95\%$ of correct predictions). However, their performance strongly decreases when predicting longer sequences composed of multiple tokens or even multiple statements.

\begin{figure*}[tb]
    \centering
    \includegraphics[width=\linewidth]{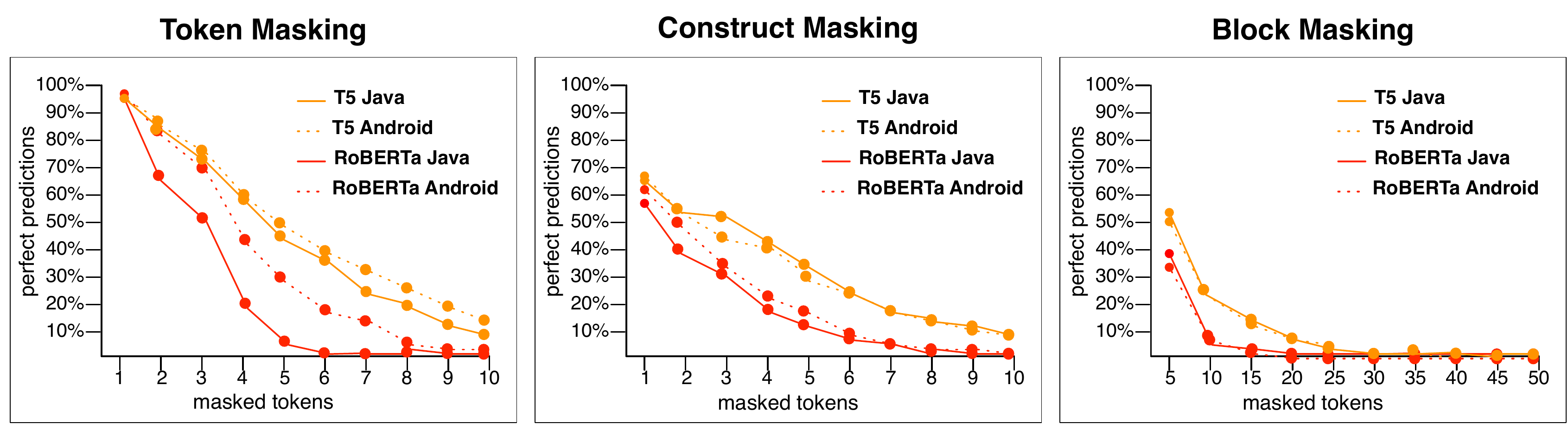}
    \caption{Percentage of perfect predictions achieved by T5 and \roberta}
    \label{fig:results}
\end{figure*}

%%%%%%%%%%%%%%%%%%%%%%%%%%%%%%%%%%%%%%%%
\subsection{DL-based models performance comparison (RQ$_1$)} \label{sec:rq1}
%%%%%%%%%%%%%%%%%%%%%%%%%%%%%%%%%%%%%%%%

\figref{fig:results} depicts the results achieved by DL-based models in terms of perfect predictions for different masking approaches, namely (from left to right) \emph{token-masking}, \emph{construct-masking}, and \emph{block-masking}. The plots show the percentage of perfect predictions ($y$ axis) by the number of masked tokens ($x$ axis). For example, in the \emph{token masking} scenario we randomly mask, for each source code line $l$ having more than one token, its last $x$ tokens, where $x$ is a random number between $1$ $\dots$ $n-1$, with $n$ being the number of tokens of $l$, and $x$ is capped to a maximum of 10. The results achieved by the T5 are reported in orange while those for \roberta in red; continuous lines represent the results achieved on the Java dataset, while the dashed lines are used for the Android dataset.

The left-side graph in \figref{fig:results} shows the percentage of perfect predictions when we only mask the last token (\ie one masked token), the last two tokens, \etc. The scale on the $x$ axis is different when dealing with the block masking scenario since here we mask entire blocks thus having, in some cases, dozens of masked tokens. Each point indicates that between $x-5$ and $x$ tokens were masked, \eg for the first data point at most 5 tokens were masked, for the second between 5 and 10, \etc.

\input{tables/bleu}

\tabref{tab:bleu} reports the average BLEU score in the four considered variants and the average normalized Levenshtein distance achieved by T5 and \roberta. Also in this case the results are grouped based on the masking level and dataset. 

\input{tables/perfect_prediction_dl_models}

The results in \figref{fig:results} and \tabref{tab:bleu} are achieved by the DL-based models in the simplest scenario, \ie single-task without pretraining. To answer RQ$_{1.3}$ we run additional experiments for the best model (\ie T5). The results of such experiments are provided in \tabref{tab:perfect_prediction_dl_models} as the percentage of perfect predictions for different variants of the T5 model, \ie with/without pretraining and using single- and multi-task fine-tuning. \tabref{tab:perfect_prediction_dl_models} also reports the results achieved with the \roberta model in the simplest scenario to simplify the discussion of the results.

%%%%%%%%%%%%%%%%%%%%%%%%%%%%%%%%%%%%%%%%
\subsubsection{Impact of number of masked tokens (RQ$_{1.1}$) and specificity of the dataset (RQ$_{1.2}$)}
%%%%%%%%%%%%%%%%%%%%%%%%%%%%%%%%%%%%%%%%
Three findings immediately emerge from the analysis of \figref{fig:results}: (i) as expected, the higher the number of masked tokens, the lower the performance of the models; (ii) the results achieved on the more specific dataset (\ie Android, dashed lines in \figref{fig:results}) are substantially better as compared to the ones achieved for Java only in the token-masking scenario with the \roberta model (see statistics in \tabref{tab:datasets-comparison}); 
(iii) the T5 model (orange lines in \figref{fig:results}) substantially outperforms \roberta (see statistics in \tabref{tab:stats-perfect} and \tabref{tab:stats-bleu}).  Also, the performance of \roberta drops more steadily as compared to that of T5 when the number of masked tokens increases.

\input{tables/stats-perfect}

\tabref{tab:stats-perfect} reports results of the McNemar's test and ORs for the comparison between T5 and \roberta in terms of their ability to perform perfect predictions. As it can be seen, the (adjusted) \emph{p}-values always indicate a statistically significant difference, and the ORs indicate that T5 has between 2.94 and 8.87 higher odds to provide a perfect prediction than \roberta.

\input{tables/stats-bleu}

Concerning the comparison of BLEU scores or Levenshtein distances (whose average values are reported in \tabref{tab:bleu}) between T5 and \roberta, statistical results (Wilcoxon signed-rank test adjusted \emph{p}-values and Cliff's $d$) are in \tabref{tab:stats-bleu}. Also in this case, differences are always statistically significant, with varying effect sizes (generally larger for greater levels of BLEU score, and for Java than Android) in favor of T5 (for the Levenshtein distance a negative $d$ is in favor of T5, as it is a distance).

\input{tables/dataset-comparison}

\textbf{Token masking.} The left part of \figref{fig:results} shows that, as expected, the lower the number of masked tokens the higher the perfect predictions. Not surprisingly, the models are very effective when we only mask the last token in a statement. Indeed, in most cases, this will be a semicolon, a parenthesis, or a curly bracket. Thus, it is easy for the model to guess the last token. When moving to more challenging scenarios like the last five tokens masked in a statement, the percentage of perfect predictions for \roberta on the Java dataset drops to less than 10\%, a major gap with the T5 model that keeps a percentage of perfect predictions higher than 40\%. As for the dataset, both models achieve significantly better performance on the Android dataset (Fisher's test p-value$<$0.001 and OR$<1$), which is more specific and, thus, more subject to regularities in the source code. However, the gap in terms of perfect predictions between the Java and the Android dataset is much more marked for the \roberta model (\eg $\sim$20\% at $x=5$ against a $\sim$6\% for the T5).

Looking at \tabref{tab:bleu}, the BLEU scores and the Levenshtein distance confirm what was observed for perfect predictions: performances for the Android dataset are better than for the Java one. According to Wilcoxon rank-sum test, all differences, except for \roberta at Block level, are statistically significant, yet with a negligible/small Cliff's $d$ (detailed statistical results are in the online appendix).

\textbf{Construct masking.} In this scenario (see central sub-graph in \figref{fig:results}), T5 and \roberta achieve respectively above 65\% and 55\% of perfect predictions when a single token is masked for both datasets. Note that, in this scenario, also a single-token prediction is not trivial since we are in a context in which such a single token represents (i) the complete condition of an {\tt if} statement or a {\tt while}/{\tt for} loop, or (ii) the parameters in a method call, or (iii) the exception caught in a {\tt catch} statement. When the prediction is represented by a single token, it is usually related to a Boolean used in an {\tt if} condition (\eg {\tt if(true)}, {\tt if(valid)}, \etc) or the single parameter needed for a method invocation. 

Also in this case, a higher number of masked tokens implies lower performance, and again the T5 outperforms \roberta significantly for both datasets although the gap is smaller. Finally, as shown in \tabref{tab:datasets-comparison}, while with \roberta results for Android are better, for T5 we achieve an OR$\simeq 1$.

In terms of BLEU score and Levenshtein distance, the achieved values are worse as compared to the token-level masking, confirming the more challenging prediction scenario represented by the construct-level masking. On average, the developer may need to modify $\sim$40\% and $\sim$30\% of the predicted tokens to obtain the reference code (small variations are observed between  Java and Android) when using \roberta and T5, respectively. 

\textbf{Block masking.} This represents the most challenging prediction scenario: The masked part can involve an entire statement or even span over two statements (maximum boundary we set). The performance of T5 and \roberta in terms of perfect predictions are respectively above 50\% and 35\% when dealing with small masked blocks, up to five tokens. These blocks are mostly related to \texttt{return} statements representing a code block (\eg the value to return when an \texttt{if} condition is satisfied), such as \texttt{ \{ return false; \}}, \texttt{ \{ return null; \}}, \etc

\begin{figure}[tb]
    \centering
    \includegraphics[width=\linewidth]{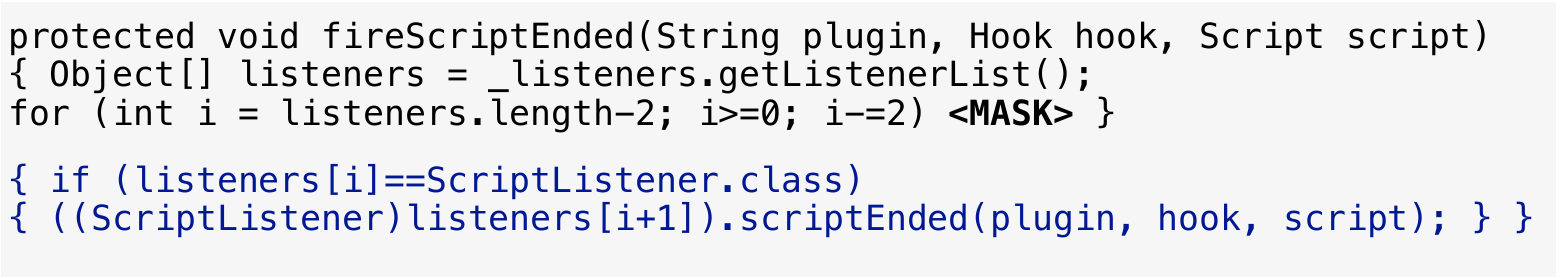}
    \caption{Perfect prediction of 36 tokens generated by T5 in the Android dataset}
    \label{fig:long_prediction}
\end{figure}

For longer blocks, the performance substantially drops. When considering blocks having between six and ten masked tokens, \roberta is able to generate a correct prediction in $\sim$5\% of cases, as compared to the $\sim$25\% achieved by the T5. The largest masked block reporting a perfect prediction for the T5 model is composed of 36 and 39 tokens for Android (see \figref{fig:long_prediction}) and Java datasets respectively, compared to the 13 and 15 tokens achieved with the \roberta model.

At this level (see \tabref{tab:datasets-comparison}), the difference in terms of performance between Java and Android is not so evident, and even insignificant for T5.

As expected, the BLEU scores are the lowest in this scenario (\tabref{tab:bleu}), and the developer may need to revise, on average, $\sim50$\% and $\sim35$\% of the predicted tokens, independently from the dataset of interest, when using \roberta and T5, respectively. \smallskip

{\bf Answer to RQ$_{1.1}$:} \emph{As the number of masked tokens increases, the DL-based models have a harder time generating correct predictions. Still, the performance achieved by the T5 model looks promising and, as we will discuss later, can be further pushed through proper pretraining and multi-task fine-tuning.}

{\bf Answer to RQ$_{1.2}$:} \emph{When looking at the best model (\ie the T5), its performance on the two datasets is quite similar, with no major differences observed. A strong difference in performance is only observed in the token-masking scenario with the \roberta model.}

%%%%%%%%%%%%%%%%%%%%%%%%%%%%%%%%%%%%%%%%
\subsubsection{Impact of pre-training and transfer learning (RQ$_{2}$)}
%%%%%%%%%%%%%%%%%%%%%%%%%%%%%%%%%%%%%%%%
As explained in \secref{sec:design:training_t5}, we trained seven additional T5 models to assess the impact of pretraining and transfer learning on its performance. First, we added to the six models for which we previously discussed the T5 performance (\ie no pretraining, single-task) the pretraining phase (obtaining a pre-trained model in the single-task scenario, \ie no transfer learning). Then, we take the pre-trained model, and fine-tuned it in a multi-task setting, investigating the impact of transfer learning. 

\input{tables/pretraining-stats}

\tabref{tab:perfect_prediction_dl_models} shows the achieved results also reporting the performance of the  previously discussed T5 and \roberta models (\ie no pretraining, single-task in \tabref{tab:perfect_prediction_dl_models}). 
Results of a statistical comparison made using McNemar's test are reported in \tabref{tab:pretraining-stats}. As it is shown, the pretraining has a positive (OR$>1$) and statistically significant effect in all cases, and the fine-tuning in a multi-task setting outperforms the single-task pretraining.
Looking at \tabref{tab:perfect_prediction_dl_models}, the pretraining had a positive impact on the accuracy of T5, boosting the percentage of perfect predictions from 1\% to 4.7\%, depending on the test dataset. The benefit of pretraining is more evident in the most challenging block-level scenario ($\sim$5\%). Overall, when considering all test datasets as a whole, the percentage of perfect predictions increases from 54.1\% to 56.2\% (+2.1\%).

By training a single model on the six training datasets, the percentage of perfect predictions further increases, going up to an overall 59.3\%. Note that improvements can be observed on all test datasets and, for the token-masking scenario, they can reach $\sim$5\%.

The performance improvement is also confirmed by the results achieved in terms of BLEU score and the Levenshtein distance that, for the sake of brevity, we report in our replication package \cite{replication}.

{\bf Answer to RQ$_{2}$:} \emph{We found both pretraining and multi-task fine-tuning to have a positive impact on the T5 performance. Overall, such an improvement accounts for +5.2\% in terms of perfect predictions (36,009 additional instances correctly predicted).}

\begin{figure}[tb]
    \centering
    \includegraphics[width=0.65\linewidth]{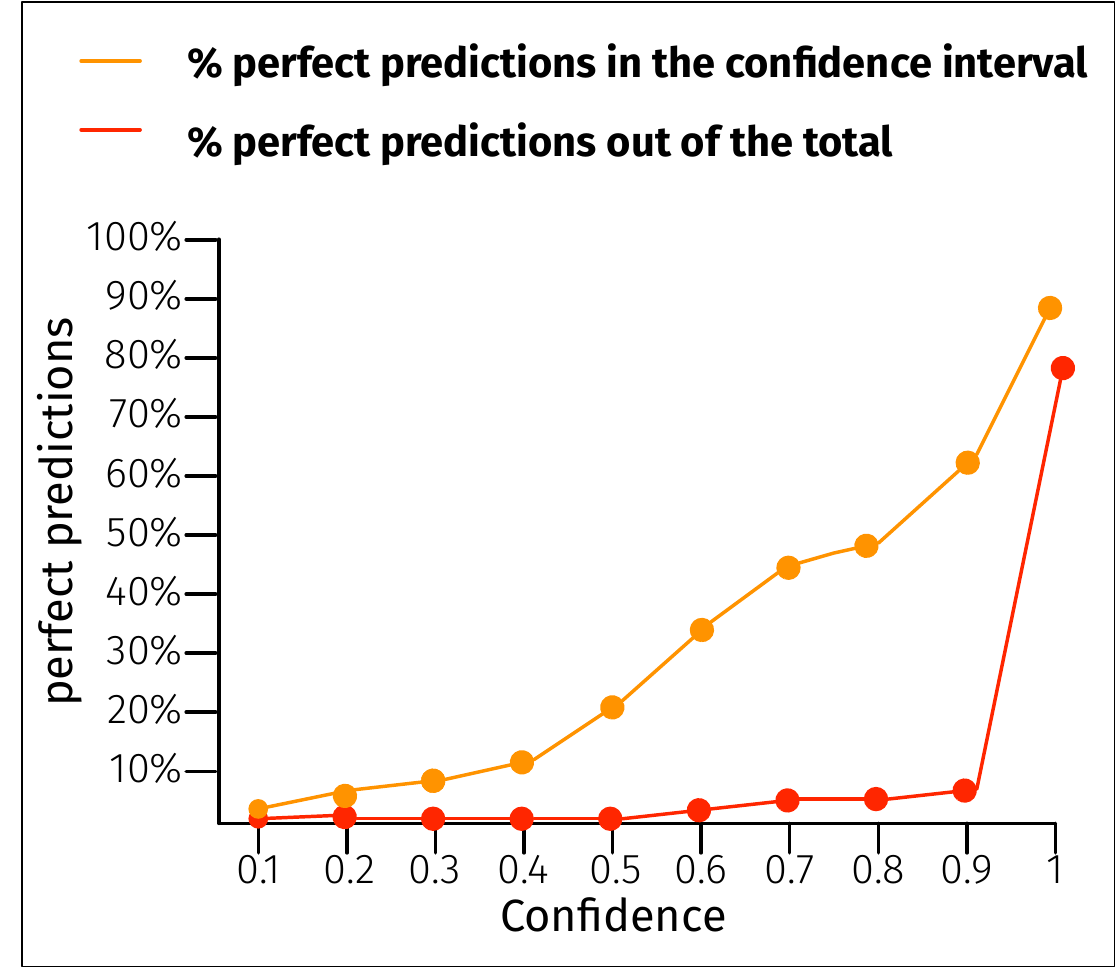}
    \caption{Perfect predictions by the confidence of the model}
    \label{fig:confidence_perfect_predictions}
\end{figure}

\subsubsection{T5 Confidence Level}
The T5 returns a \emph{score} for each prediction, ranging from minus infinity to 0. This score is the log-likelihood of the prediction itself. If the score is -2 then it means that the log-likelihood of the prediction is -2. Hence, the likelihood is 0.14 ($ln(x)=-2 \Longrightarrow x=0.14$) and this implies that the model has a confidence of 14\% for the prediction to be correct. If the score is 0, repeating the same computation as above, the model has the confidence of 100\% about the prediction itself.

\figref{fig:confidence_perfect_predictions} reports the relationship between the percentage of perfect predictions and the confidence of the model. The orange line shows the percentage of perfect predictions within each confidence interval (\eg 90\% of predictions having a confidence higher than 0.9 are correct), while the red line reports the percentage of perfect predictions that are due to predictions in that confidence interval out of the total (\eg 78\% of all perfect predictions have a confidence higher than 0.9).

\figref{fig:confidence_perfect_predictions} shows a strong relationship between the confidence of the model and the correctness of the prediction. While this result might look minor, it has an important implication: It would be possible to build a reliable code completion tool around the T5 model. Indeed, the tool could be configured to only trigger recommendations when the confidence of the prediction is higher than a given threshold (\eg 0.9). This would result in an extremely high precision. 

From a statistical perspective, a logistic regression model
 correlating the confidence level and the perfect prediction outcome indicates a statistically significant (\emph{p}-value $<$0.001) correlation, and an estimate of 6.58, which means 720 higher odds of a perfect prediction for each unit increase of the confidence, \ie 72 higher odds of a perfect prediction for a 0.1 increase of the confidence, \ie a tick on the x-axis of \figref{fig:confidence_perfect_predictions}. 

\begin{figure}[tb]
    \centering
    \includegraphics[width=0.68\linewidth]{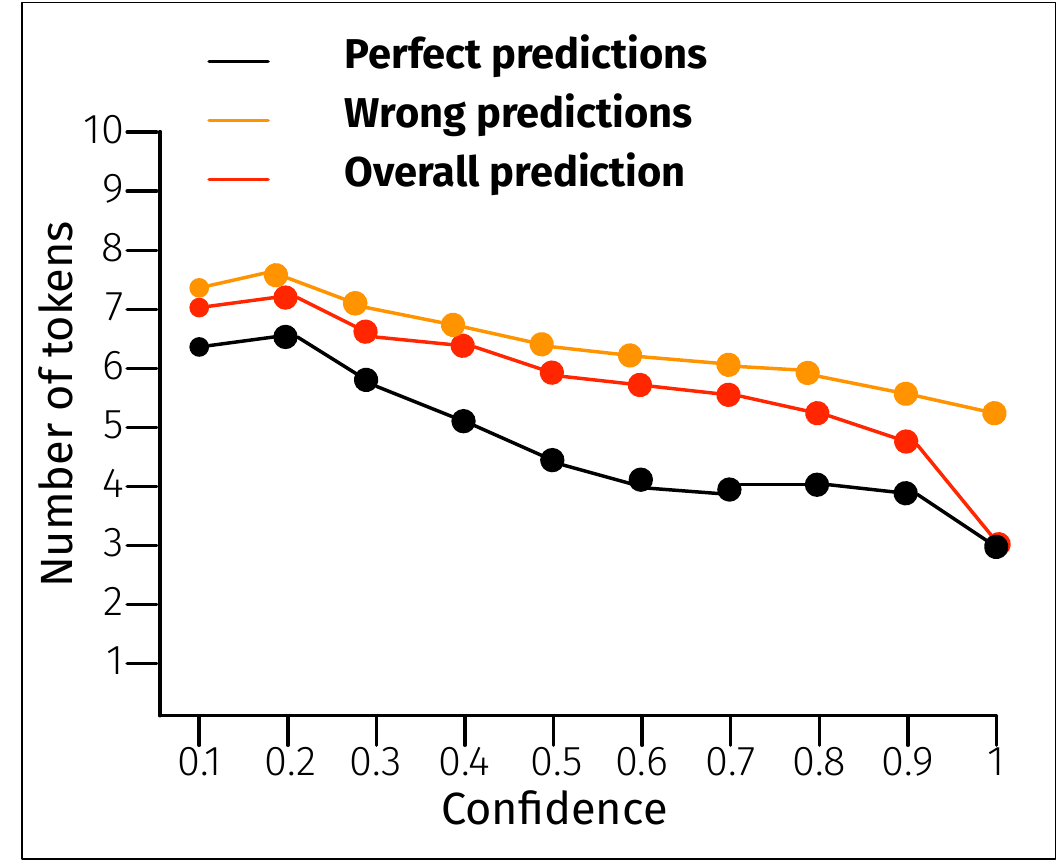}
    \caption{Average length (in tokens) of the predictions by confidence}
    \label{fig:prediction_length}
\end{figure}

\figref{fig:prediction_length} analyzes the average length, in tokens, of the perfect predictions (yellow line), wrong predictions (orange line), and for all the predictions (red line) among all confidence intervals. It is clear that the length of the prediction is related to the confidence, since the model has higher confidence for shorter predictions. Indeed, the average number of tokens in perfect predictions for the highest confidence interval (\ie 3 tokens) is much lower than the average number of tokens in perfect predictions for the lowest confidence interval (\ie 6 tokens). This confirms previous findings showing that the model is more likely to correctly predict shorter statements.

From a statistical perspective, this is confirmed by a significant (\emph{p}-value $<$0.001), negative, and moderate Kendall's correlation ($\tau$=-0.36).

\input{tables/ngram} 
\input{tables/stats-comparison}
%%%%%%%%%%%%%%%%%%%%%%%%%%%%%%%%%%%%%%%%
\subsection{Comparison with an n-gram Model}  \label{sec:rq2}
%%%%%%%%%%%%%%%%%%%%%%%%%%%%%%%%%%%%%%%%
We answer RQ$_3$ by comparing the DL-based models without pretraining and in the single-task setting to the $n$-gram model. We opted for this comparison for the sake of fairness, since in this way the $n$-gram model has been trained on exactly the same dataset as the two DL-based models.

\tabref{tab:perfect_prediction_vs_ngram} reports the comparison in terms of perfect predictions between T5, \roberta and the $n$-gram model in different evaluation scenarios, as well as the overall results. For example, T5 produced 61\% perfect predictions on the Java dataset when using token masking. Results of statistical tests (McNemar's test) are in \tabref{tab:comparison}.

One important clarification is needed to properly interpret the results of \tabref{tab:perfect_prediction_vs_ngram}. Since the $n$-gram model uses a different script to tokenize the code, we excluded from the test sets cases in which the tokens to predict (\ie the masked ones) are tokenized differently between the DL-based approaches and $n$-gram one (\eg one identifies 4 tokens and the other one 5). This resulted in the exclusions of a few hundred instances from each test set and explains the slightly different performances reported for T5 and \roberta between \tabref{tab:perfect_prediction_vs_ngram} and \figref{fig:results}. 

\tabref{tab:comparison} reports results of the statistical comparison among the three models, using McNemar's test.
DL-based models achieve better performance in all experimented datasets, and McNemar's tests always indicate statistically significant differences, with ORs ranging between 1.90 (\roberta vs n-grams, block masking for Android) and 14.38 (block masking, T5 vs n-grams for Java).

In the token masking scenario, the performance of the $n$-gram model is very competitive when compared with \roberta, while the T5 performs substantially better. When masking specific constructs, the gap in performance becomes stronger (see \tabref{tab:perfect_prediction_vs_ngram}) with a substantial gap, especially between T5 and $n$-gram. Finally, in the block masking experiment, \roberta and $n$-gram techniques struggle to obtain a high percentage of perfect predictions, with the T5 performing better achieving more than twice the number of perfect predictions as compared to the competitive techniques.

While the DL-based models showed superior performance, there are two important aspects to consider. First, the $n$-gram model allows for faster training. We estimate four to five times less training time needed for the $n$-gram model as compared to the DL-based models. We do not report precise data since such a study would require executing the training many times on the same machine, and such an analysis is out of the scope of this work. Once trained all models can generate predictions in fractions of a second. Second, the comparison presented as of now concerns the standard $n$-gram model. However, we also experimented with the cached $n$-gram model \cite{Hellendoorn:fse2017}, which can leverage information about other code components coming from the same project (\eg same file or package \cite{Hellendoorn:fse2017}) of the method in which the prediction is performed. This is one of the advantages of the cache model \cite{Hellendoorn:fse2017} and, in a real scenario, it should be possible to use this information assuming that the method on which the prediction is performed is not the first one written in the whole system. However, such experimentation is quite expensive to perform since it requires the cloning of the whole repositories hosting every test method. This is why it has only been performed on a small sample of our dataset.  

\input{tables/ngram_cloned}

For a given method $m_t$ in the test set, we clone its repository and check if the source code of $m_t$ in the latest system snapshot is exactly the same as in the test set. If this is the case, we run the prediction on $m_t$ providing the cloned repository as a test folder, in such a way that it is leveraged by the cache model (this is done through the implementation of Hellendoorn \etal \cite{Hellendoorn:fse2017}).
If the method changed, we discard it and move to the next one. Since such a process is very expensive, we collected 200 methods from each test set, and we compare the performance of the $n$-gram model when such additional information is provided (and not) on these instances. 

\tabref{tab:perfect_prediction_vs_ngram_cloned} reports the achieved results. As expected, the performance of the $n$-gram model increase thanks to the use of the information in the test project. On these same instances, the performance of T5 and \roberta models are always superior but in the case of Java token and block masking for \roberta.

{\bf Answer to RQ$_3$:} \emph{The $n$-gram model is a competitive alternative to \roberta, while the T5 confirms its superior performance. It is worth highlighting the much cheaper cost of training (and possibly re-training several times) an $n$-gram model as compared to a DL-based approach.}

\subsection{Qualitative Results}  \label{sec:rq1_qualitative}

\begin{figure}[tb]
    \centering
    \includegraphics[width=0.95\linewidth]{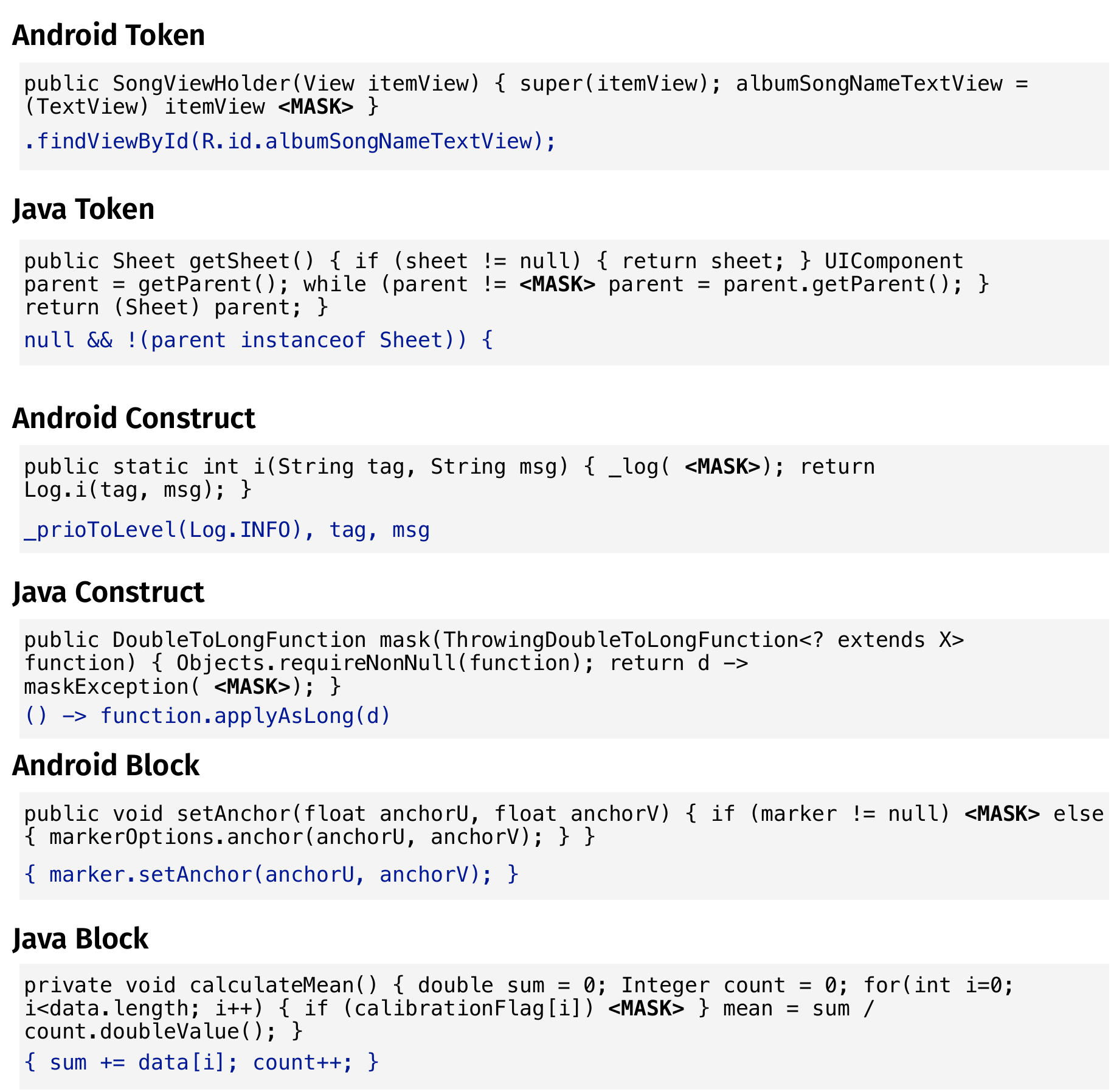}
    \caption{Examples of perfect predictions generated by T5}
    \label{fig:qualitative_T5}
\end{figure}

To give a better idea to the reader about the capabilities of the experimented models in supporting code completion, we report in \figref{fig:qualitative_T5} examples of correct predictions for the T5 model in different scenarios/datasets. Examples of predictions for the \roberta and $n$-gram model are available in the replication package \cite{replication}.

Given the achieved results showing the superiority of the T5 model, we had a better look at a sample of the wrong predictions it generates, to see whether some of them are semantically correct (\eg \emph{return 0x0;} is equivalent to \emph{return 0;}) despite being different from the reference code written by the developers. The first author looked at 200 wrong predictions generated within the highest confidence interval, finding that only in three cases the prediction was semantically equivalent, with the reference code including extra (unnecessary) brackets not generated by the T5 model (\eg T5 predicts \emph{entry;} instead of \emph{(entry);}). Overall, it appeared that several of the generated predictions, while wrong, might still speed up the implementation process, for example when $n-1$ out of the $n$ parameters needed for a method invocation are correctly predicted. Clearly, only a user study with developers can help in assessing the actual usefulness of these predictions during real coding activities.

Since we found cases in which the perfect predictions of the T5 spanned across dozens of tokens, being almost unrealistic, we checked whether the 21 perfect predictions having more than 30 tokens were already present in the training set. Indeed, while we ensure that there are no duplicated methods between training and test, it is possible that two different methods $m_1$ and $m_2$ have the same masked part (\ie the two methods are different in the non-masked part but they have the same set of masked tokens). Only one out of the 21 inspected cases was already present in the training set and related to the transpose of a matrix. The model was able to correctly predict very complex masked parts such as \emph{"\{ if (defaultProviders != null  \&\& index < defaultProviders.length) \{ return defaultProviders[index].getRebuild(defaultProviders, index + 1); \} \} "}.

Finally, it is worth commenting on the possible reasons behind the superior performance we observed for the T5 as compared to \roberta and for the DL-based models as compared to the $n$-gram model. \roberta predicts all of the masked tokens at the same time, whereas T5 predicts them one by one. This means that \roberta cannot use the previously generated tokens to predict the next one, while the T5 exploits this additional information. Concerning the superior performance of the DL-based model as compared to the $n$-grams, this most likely comes down to the context window it is able to see. Indeed, the $n$-gram model can only see (and leverage) a few tokens when predicting the next one, while both T5 and \roberta have a better view of the coding context, seeing all the tokens surrounding the masked ones (which could be hundreds). A solution could be to scale up the $n$-gram model which, however, would become too demanding in terms of computational cost.

%% file: tables/bleu.tex
% !TEX root = ../main.tex

% Design tables. 

\begin{table}[h]
	\centering
	\caption{BLEU score and Levenshtein distance for T5 and \roberta.}
	\scriptsize
	\label{tab:bleu}
	\begin{tabular}{lrrrrr}
	\toprule
	\multicolumn{6}{c}{{\bf Token masking}}\\\midrule
	& \multicolumn{2}{c}{{\bf Java}} & & \multicolumn{2}{c}{{\bf Android}}\\ \cline{2-3} \cline{5-6}
	& {\bf T5} & {\bf RoBERTa} & & {\bf T5} & {\bf RoBERTa}\\\hline
	BLEU-1 & 0.83 & 0.60 && 0.85 & 0.73\\
	BLEU-2 & 0.73 & 0.43 && 0.76 & 0.61\\
	BLEU-3 & 0.60 & 0.23 && 0.64 & 0.44\\
	BLEU-4 & 0.47 & 0.10 && 0.51 & 0.28\\
	Levenshtein & 0.16 & 0.35 && 0.14 & 0.24\\\midrule

	\multicolumn{6}{c}{{\bf Construct masking}}\\\midrule
	& \multicolumn{2}{c}{{\bf Java}} & & \multicolumn{2}{c}{{\bf Android}}\\ \cline{2-3} \cline{5-6}
	& {\bf T5} & {\bf RoBERTa} & & {\bf T5} & {\bf RoBERTa}\\\hline
	BLEU-1 & 0.68 & 0.51 && 0.68 & 0.57\\
	BLEU-2 & 0.55 & 0.34 && 0.57 & 0.43\\
	BLEU-3 & 0.48 & 0.24 && 0.49 & 0.33\\
	BLEU-4 & 0.37 &0.14 && 0.43 & 0.26\\
	Levenshtein & 0.32 & 0.48 && 0.32 & 0.41 \\\midrule
	
	\multicolumn{6}{c}{{\bf Block masking}}\\\midrule
	& \multicolumn{2}{c}{{\bf Java}} & & \multicolumn{2}{c}{{\bf Android}}\\ \cline{2-3} \cline{5-6}
	& {\bf T5} & {\bf RoBERTa} & & {\bf T5} & {\bf RoBERTa}\\\hline
	BLEU-1 & 0.65 & 0.44 && 0.62 & 0.44\\
	BLEU-2 & 0.57 & 0.32 && 0.54 & 0.31\\
	BLEU-3 & 0.49 & 0.21 && 0.46 & 0.21\\
	BLEU-4 & 0.41 & 0.13 && 0.38 & 0.13\\
	Levenshtein & 0.35 & 0.54 &&0.37 & 0.55\\\bottomrule
\end{tabular} 
\end{table}

%% file: tables/perfect_prediction_dl_models.tex
% !TEX root = ../main.tex

\begin{table*}
	\centering
	\caption{Perfect predictions of T5 models with different fine-tuning strategies, and \roberta model}
	\scriptsize
	\label{tab:perfect_prediction_dl_models}
	\begin{tabularx}{.7\textwidth}{llrrrrrr}
	\toprule
	\multicolumn{2}{c}{\multirow{3}{*}{{\bf Dataset and Masking Level}}} &\multicolumn{4}{c}{{\bf T5}}  & & \multicolumn{1}{c}{{\bf RoBERTa}}\\ \cline{3-6} \cline{8-8}
	&&  \multicolumn{2}{c}{{\bf With Pretraining}} && {\bf No Pretraining} & & {\bf No Pretraining}\\ \cline{3-4} \cline{6-6} \cline{8-8}
		&& {\bf Single-task} & {\bf Multi-task} && {\bf Single-task} & & {\bf Single-task}\\\midrule
	
	\multirow{3}{*}{Java}		&Token&  62.9\% & 66.3\% && 61.0\% & & 38.9\% \\
				    		&Construct&  51.2\% & 53.0\% && 48.4\% & & 33.4\%  \\
				    		&Block&  27.2\% & 28.8\% && 22.9\% & &8.7\%  \\ \midrule
	\multirow{3}{*}{Android}	&Token& 64.8\% & 69.3\% &&  63.8\% & &51.8\%  \\
						&Construct& 49.3\% & 50.8\% && 46.8\% & & 37.4\%  \\
						&Block& 27.5\% & 29.7\% && 22.8\% & &9.4\% \\ \midrule
	\multicolumn{2}{c}{ {\bf Overall} } & 56.2\% & 59.3\% && 54.1\% & & 38.7\%\\
	
\bottomrule
\end{tabularx} 
\vspace{-0.2cm}
\end{table*}

%% file: tables/stats-perfect.tex
\begin{table}[ht]
	\centering
	\caption{Perfect prediction: Mcnamar's test comparison between T5 and \roberta}
        \label{tab:stats-perfect}
	\begin{tabular}{llrr}
		\hline
		\textbf{Dataset} & \textbf{Masking} & \textbf{\emph{p}-value} & \textbf{OR} \\ 
		\hline
		\multirow{3}{*}{Java} & Token & $<$0.001 & 8.87 \\ 
		& Construct & $<$0.001 & 4.69 \\ 
		& Block & $<$0.001 & 8.14 \\
		\multirow{3}{*}{Android} & Token & $<$0.001 & 4.47 \\ 
		 & Construct & $<$0.001 & 2.94 \\ 
		 & Block & $<$0.001 & 7.61 \\ 
		\hline
	\end{tabular}
\end{table}

%% file: tables/stats-bleu.tex
\begin{table*}[ht]
\centering
\caption{BLEU score and Levensthein distance comparison between T5 and \roberta: Wilcoxon signed-rank and Cliff's delta (N: negligible, S: small, M: medium, L: large)\vspace{-0.3cm}}
\label{tab:stats-bleu}
\begin{tabular}{llrrrrrrrrrr}
  \hline
\multirow{2}{*}{\textbf{Dataset}} & \multirow{2}{*}{\textbf{Masking}} & \multicolumn{2}{c}{BLEU 1}
& \multicolumn{2}{c}{BLEU 2} & \multicolumn{2}{c}{BLEU 3} &  \multicolumn{2}{c}{BLEU 4} &  \multicolumn{2}{c}{Levenshtein}\\
 & & \textbf{\emph{p}-value} & \textbf{\emph{d}} 
 & \textbf{\emph{p}-value} & \textbf{\emph{d}} 
 & \textbf{\emph{p}-value} & \textbf{\emph{d}} 
 & \textbf{\emph{p}-value} & \textbf{\emph{d}} 
 & \textbf{\emph{p}value} & \textbf{\emph{d}} \\
  \hline
\multirow{3}{*}{Java} & Token & $<$0.001 & 0.33 (S) & $<$0.001 & 0.41 (M) & $<$0.001 & 0.51 (L) & $<$0.001 & 0.62 (L) & $<$0.001 & -0.32 (S) \\
   & Construct & $<$0.001 & 0.22 (S) & $<$0.001 & 0.30 (S) & $<$0.001 & 0.32 (S) & $<$0.001 & 0.35 (M) & $<$0.001 & -0.21 (S) \\
   & Block & $<$0.001 & 0.39 (M) & $<$0.001 & 0.43 (M) & $<$0.001 & 0.47 (M) & $<$0.001 & 0.49 (L) & $<$0.001 & -0.38 (M) \\ \hline
  \multirow{3}{*}{Android} & Token & $<$0.001 & 0.17 (S) & $<$0.001 & 0.21 (S) & $<$0.001 & 0.27 (S) & $<$0.001 & 0.34 (M) & $<$0.001 & -0.17 (S) \\
   & Construct & $<$0.001 & 0.14 (N) & $<$0.001 & 0.20 (S) & $<$0.001 & 0.22 (S) & $<$0.001 & 0.27 (S) & $<$0.001 & -0.14 (N) \\
   & Block & $<$0.001 & 0.33 (M) & $<$0.001 & 0.39 (M) & $<$0.001 & 0.42 (M) & $<$0.001 & 0.44 (M) & $<$0.001 & -0.34 (M) \\
   \hline
\end{tabular}
\end{table*}

%% file: tables/dataset-comparison.tex
\begin{table}[ht]
\centering
\caption{Comparison between different datasets for perfect predictions - results of Fisher's exact test (OR$<$1 indicate better performances for Android)}
\label{tab:datasets-comparison}
\begin{tabular}{llrr}
  \hline
\textbf{Masking} & \textbf{Method}& \textbf{\emph{p}-value} & \textbf{OR} \\ 
  \hline
\multirow{2}{*}{Token} & T5 & $<$0.001 & 0.89 \\ 
   & RoBERTa & $<$0.001 & 0.59 \\ \hline
  \multirow{2}{*}{Construct} & T5 & $<$0.001 & 1.07 \\ 
  & RoBERTa & $<$0.001 & 0.84 \\ \hline
  \multirow{2}{*}{Block} & T5 & 0.67 & 1.01 \\ 
   & RoBERTa & 0.01 & 0.93 \\ 
   \hline
\end{tabular}
\end{table}

%% file: tables/pretraining-stats.tex
\begin{table}[ht]
\centering
\caption{Effect of different pretraining levels for T5: McNemar's test results. None indicates the T5 model with no pre-training and single-task finetuning. Single and Multi indicates the pre-trained model with single- and multi-task fine-tuning, respectively.}
\label{tab:pretraining-stats}
\begin{tabular}{lllrr}
  \hline
\textbf{Dataset} & \textbf{Masking} & \textbf{Comparison} & \textbf{\emph{p}-value} & \textbf{OR} \\ 
  \hline
        \multirow{9}{*}{Java}   & \multirow{3}{*}{Token} & single vs. none & $<$0.001 & 1.44 \\ 
  &  & multi vs. single & $<$0.001 & 1.81 \\ 
  &  & multi vs none & $<$0.001 & 2.33 \\ 
  \cline{2-5}
  & \multirow{3}{*}{Construct} & single vs. none & $<$0.001 & 1.61 \\ 
  &  & multi vs. single & $<$0.001 & 1.34 \\ 
  &  & multi vs none & $<$0.001 & 1.92 \\ 
  \cline{2-5}
  & \multirow{3}{*}{Block} & single vs. none & $<$0.001 & 2.19 \\ 
  &  & multi vs. single & $<$0.001 & 1.32 \\ 
  &  & multi vs none & $<$0.001 & 2.32 \\ 
  \hline
   \multirow{9}{*}{Android} & \multirow{3}{*}{Token} & single vs. none & $<$0.001 & 1.23 \\ 
   &  & multi vs. single & $<$0.001 & 2.27 \\ 
   &  & multi vs none & $<$0.001 & 2.61 \\ 
   \cline{2-5}
   & \multirow{3}{*}{Construct}  & single vs. none & $<$0.001 & 1.58 \\ 
   &  & multi vs. single & $<$0.001 & 1.28 \\ 
   &  & multi vs none & $<$0.001 & 1.81 \\ 
      \cline{2-5}
   & \multirow{3}{*}{Block} & single vs. none & $<$0.001 & 2.14 \\ 
   &  & multi vs. single & $<$0.001 & 1.39 \\ 
   &  & multi vs none & $<$0.001 & 2.39 \\ 
  \hline
\end{tabular}
\end{table}

%% file: tables/ngram.tex
% !TEX root = ../main.tex

\begin{table}[h]
	\centering
	\caption{Perfect predictions of the three models}
	\scriptsize
	\label{tab:perfect_prediction_vs_ngram}
	\begin{tabularx}{.4\textwidth}{llrrr}
	\toprule
	\multicolumn{2}{c}{{\bf Dataset and Masking Level}} & {\bf T5} & {\bf RoBERTa}  & {\bf $n$-gram} \\  \midrule
		
	\multirow{3}{*}{Java}		&Token		&  61.0\%	& 38.9\% 	& 30.4\% \\
				    		&Construct	&  48.8\% 	& 33.9\%	& 12.5\% \\
				    		&Block		&  22.9\% 	& 8.7\% 	& 4.6\%   \\ \midrule
	\multirow{3}{*}{Android}	&Token		& 63.8\% 	&51.9\% 	& 35.4\% \\
						&Construct	& 47.1\% 	& 37.8\%  	& 17.6\% \\
						&Block		& 22.8\%	&9.4\%	 & 6.6\% \\ \midrule
%    	Java token masking &  62.9\% & 66.3\% & 61.0\% & & 38.9\%\\
%    	Java construct masking & 51.2\% & 53.0\% & 48.4\% & & 33.4\%\\	
%    	Java block masking & 27.2\% & 28.8\% & 22.9\% & &8.7\%\\
%    	Android token masking & 64.8\% & 69.3\% &  63.8\% & &51.8\%\\
%    	Android construct masking & 49.3\% & 50.8\% & 46.8\% & & 37.7\%\\
%    	Android block masking & 27.5\% & 29.7\% & 22.8\% & &9.3\%\\
	\multicolumn{2}{c}{ {\bf Overall} } & 54.3\% & 38.8 & 24.9\%\\
	
\bottomrule
\end{tabularx} 
\end{table}

%% file: tables/stats-comparison.tex
\begin{table}[ht]
\centering
\caption{Comparison with the n-grams model: results of McNemar's test}
\label{tab:comparison}
\begin{tabular}{lllrr}
  \hline
\textbf{Dataset} & \textbf{Masking} & \textbf{Comparison} & \textbf{\emph{p}-value} & \textbf{OR} \\ 
  \hline
 \multirow{9}{*}{Java} & \multirow{3}{*}{Token} & T5 vs. RoBERTa & $<$0.001 & 8.93 \\ 
&  & RoBERTa vs. n-grams & $<$0.001 & 2.21 \\ 
&  & T5 vs. n-grams & $<$0.001 & 10.31 \\ 
\cline{2-5}
& \multirow{3}{*}{Construct} & T5 vs. RoBERTa & $<$0.001 & 4.65 \\ 
&  & RoBERTa vs. n-grams & $<$0.001 & 5.29 \\ 
&  & T5 vs. n-grams & $<$0.001 & 11.62 \\ 
\cline{2-5}
& \multirow{3}{*}{Block} & T5 vs. RoBERTa & $<$0.001 & 8.15 \\ 
&  & RoBERTa vs. n-grams & $<$0.001 & 2.85 \\ 
& & T5 vs. n-grams & $<$0.001 & 14.38 \\ 
\hline
\multirow{9}{*}{Android} & \multirow{3}{*}{Token} & T5 vs. RoBERTa & $<$0.001 & 4.47 \\ 
   &  & RoBERTa vs. n-grams & $<$0.001 & 4.26 \\ 
   &  & T5 vs. n-grams & $<$0.001 & 10.14 \\ 
   \cline{2-5}
   & \multirow{3}{*}{Construct} & T5 vs. RoBERTa & $<$0.001 & 2.91 \\ 
   &  & RoBERTa vs. n-grams & $<$0.001 & 5.30 \\ 
   &  & T5 vs. n-grams & $<$0.001 & 9.04 \\ 
   \cline{2-5}
   & \multirow{3}{*}{Block} & T5 vs. RoBERTa & $<$0.001 & 7.62 \\ 
   &  & RoBERTa vs. n-grams & $<$0.001 & 1.90 \\ 
   &  & T5 vs. n-grams & $<$0.001 & 10.00 \\ 
   \hline
\end{tabular}
\end{table}

%% file: tables/ngram_cloned.tex
% !TEX root = ../main.tex

\begin{table}[h]
	\centering
	\caption{Perfect predictions of $n$-gram model when providing the cloned repository (WC) vs. when not providing (NC). In comparison to  DL-based models (200 methods)}
	\scriptsize
	\label{tab:perfect_prediction_vs_ngram_cloned}
	\begin{tabularx}{.45\textwidth}{llrrrr}
	\toprule
	\multicolumn{2}{c}{\multirow{2}{*}{{\bf Dataset and Masking Level}}} & \multirow{2}{*}{{\bf T5}} & \multirow{2}{*}{{\bf RoBERTa}}  & \multicolumn{2}{c}{{\bf $n$-gram}} \\  \cline{5-6}
	 					&			&		& 		& NC		& WC \\ \midrule
		
	\multirow{3}{*}{Java}		&Token		&  65.5\%	& 42.2\% 	& 32.5\% 	& 43.9\% \\
				    		&Construct	&  56.0\% 	& 38.0\%	& 14.5\% 	& 20.5\% \\
				    		&Block		&  25.8\% 	& 8.5\% 	& 5.2\% 	& 8.5\% \\ \midrule
	\multirow{3}{*}{Android}	&Token		& 69.9\% 	&50.9\% 	& 35\% 	& 42.2\% \\
						&Construct	& 52.8\% 	& 37.8\%  	& 13.9\% 	& 22.0\% \\
						&Block		& 33.6\%	&13.0\%	 & 9\% 	& 11.9\% \\ \midrule
	\multicolumn{2}{c}{ {\bf Overall} } 		& 57.7\%  & 38.2\%	 & 23.9\%  	& 31.5\% \\
	
\bottomrule
\end{tabularx} 
\end{table}

%% file: threats.tex
% !TEX root = main.tex
%%%%%%%%%%%%%%%%%%%%%%%%%%%%%%%%%%%%%%%%
%%%%%%%%%%%%%%%%%%%%%%%%%%%%%%%%%%%%%%%%
\section{Threats to Validity} \label{sec:threats}
%%%%%%%%%%%%%%%%%%%%%%%%%%%%%%%%%%%%%%%%
%%%%%%%%%%%%%%%%%%%%%%%%%%%%%%%%%%%%%%%%

Threats to \emph{construct validity} concern the relationship between theory and observation. One threat, also discussed by Hellendoorn \etal \cite{HellendoornPGB19}, is related to how we simulate the extent to which code completion intervenes during development, \ie by masking source code elements. As explained in \secref{sec:datasets}, we consider different masking levels, not only to evaluate the amount of code completion that can be predicted but also to simulate different ways a developer writes source code, especially because we cannot assume this is done sequentially. However, we are aware that the considered masking levels cover a limited number of cases that may not completely reflect how developers write code.

Another threat is related to how we assess the code completion performances. On the one hand, 100\% BLEU score clearly reflects a perfect prediction. However, the BLEU score may be sufficient to assess the performance of code-related tasks \cite{Ren:codebleu} and, in general, it is difficult to evaluate the usefulness of semantic equivalent predictions or imperfect yet useful. To mitigate this threat, we report some qualitative examples, indicating how partially-complete recommendations could still be useful.

Threats to \emph{internal validity} concern factors, internal to our study, that could influence its results. To this extent, an important factor that influences DL performance is the calibration of hyperparameters, which has been performed as detailed in \secref{sec:analysis}. We are aware that due to feasibility reasons we only performed a limited calibration of the hyperparameters. Hence, it is possible that a more detailed calibration would produce better performances. Also, note that we did not experiment with a pre-trained version of \roberta. Indeed, to simplify our experimental design and reduce the training cost we decided to only pre-train the best-performing model (\ie T5).

When building the pre-training dataset we capped to 1,500 the maximum number of instances that a single project can contribute to our dataset. This has been done to avoid a handful of projects strongly influencing the training of the model. We acknowledge that different (and maybe better) results could be obtained by considering the whole code base of each project for pre-training.

Threats to \emph{conclusion validity} concern the relationship between evaluation and outcome. As explained in \secref{sec:analysis} we used appropriate statistical procedures, also adopting \emph{p}-value adjustment when multiple tests were used within the same analysis.

Threats to  \emph{external validity} are related to the generalizability of our findings. On the one hand, we have evaluated the performances of the models on two large datasets. At the same time, we do not know whether the obtained results generalize to different domains than Android, and other programming languages than Java.
A further threat is that our study is limited to the \roberta and T5 models for DL and, as a baseline for \textit{n}-gram models, the one by Hellendoorn and Devanbu~\cite{Hellendoorn:fse2017}. While we claim such models are well-representative of the current state-of-the-art, it would be desirable to investigate how alternative approaches would work for the different evaluation scenarios. Also, when building our fine-tuning dataset, we started from the CodeSearchNet Java Dataset provided by Husain \etal \cite{Java:CodeSearchNet}. In this dataset, short methods (those having less than three lines), as well as methods containing \emph{test} in their name have been excluded. This means that the results of our study do not generalize, for example, to very short methods implementing critical tasks in less than three lines of code.

%% file: related.tex
% !TEX root = main.tex
 
%%%%%%%%%%%%%%%%%%%%%%%%%%%%%%%%%%%%%%%%
%%%%%%%%%%%%%%%%%%%%%%%%%%%%%%%%%%%%%%%%
\section{Related Work} \label{sec:related}
%%%%%%%%%%%%%%%%%%%%%%%%%%%%%%%%%%%%%%%%
%%%%%%%%%%%%%%%%%%%%%%%%%%%%%%%%%%%%%%%%

We start by detailing the literature related to code completion techniques and, more specifically, we highlight the approaches aimed at (partially) automating code writing. Then, we present studies investigating the effectiveness of code completion techniques. For the sake of brevity, we do not discuss recently proposed techniques for automating bug-fixing \cite{Tufano:tosem2019,Chen:2019,Bader:oopsla2019}, modeling activities \cite{mader2019reactive}, learning code changes \cite{Tufano:icse2019,brody2020neural}, as well as source code search engines that can be used to identify pieces of code for reuse \cite{Bajracharya2006,Reiss2009,Thummalapenta2007b,Thummalapenta2008,Grechanik2010,McMillan2012}.  

\subsection{Code Completion Approaches}

The Prospector tool by Mandelin \etal \cite{MandelinXBK05} is one of the first techniques aimed at supporting code completion by suggesting within the IDE variables or method calls from the user's code base. Prospector was then followed by improvements such as the InSynth tool by Gvero \etal \cite{GveroKKP13} which, given a type expected at a given point in the source code, searches for type-compatible expressions. Other approaches focus on specific elements of API usage completion. The work from Zhang \etal~\cite{ZhangYZFZZO12} aims at recommending parameter usages, achieving 64\% of useful recommendations and 53\% of perfect ones.

Hill and Rideout \cite{Hill:2004} proposed a technique to automatically complete the body of a method. Their approach can support such a completion  for what the authors define as ``atomic clones'' (\ie small units of implementation that are unavoidable in Java to implement specific requirements). The presented tool uses the K-Nearest Neighbour to identify a clone of a method under development. Such a clone is then used to recommend the completion of the method body.

Bruch \etal~\cite{Bruch:fse2009} introduced the intelligent code completion system, able to filter out from the list of candidate method calls recommended by the IDE those that are more relevant to the current working context. 
Their results show the capability to correctly predict up to 82\% of method calls actually needed by developers, and up to 72\% of those that are relevant to the current development context. 
The approach by Bruch \etal has been improved by Proksch \etal~\cite{ProkschLM15}, by adding further contextual information and by proposing a Pattern-based Bayesian Networks approach. As a result, Proksch \etal were able to substantially reduce the model size while keeping about the same level of prediction accuracy. Differently from the aforementioned approaches, we do not restrict code completion to method calls.

Han \etal \cite{Han:2009} proposed a technique exploiting a Hidden Markov Model (HMM) to autocomplete multiple keywords starting from abbreviated inputs. This means that the user (\ie the developer) only writes a few characters of the keyword of interest that is then expanded by the HMM. The authors show that their model can save up to 41\% of keystrokes.

Robbes and Lanza~\cite{Robb2010a} used information extracted from the change history of software systems to support the code completion of method calls and class names. Their approach has been implemented in a tool named OCompletion, and the performed empirical evaluation demonstrated its ability to propose a correct match in the top-3 results in 75\% of cases.

Asaduzzaman \etal~\cite{Asaduzzaman2014} proposed a technique named CSCC (Context Sensitive Code Completion). They collect code examples from software repositories and, for each method call, represent its context as a set of methods, keywords, class, and interface names appearing within four lines of code. This contextual information is then used to filter out method call recommendations. The assumption is that similar contexts imply similar method calls. CSCC outperforms previous approaches, achieving 86\% precision and 99\% recall.

Hindle \etal~\cite{Hindle:icse2012} pioneered the work on statistical language models applied to software. They conceived the idea of ``naturalness of source code'' and used n-gram models to create a language-agnostic algorithm that is able to predict the next token in a given statement. The trained model's average entropy is between three and four bits, indicating a high degree of naturalness.

Raychev \etal~\cite{Raychev:pldi14} approach the code completion problem through statistical language models. They extract sequences of method calls from a large code base, and use this dataset to train a language model able to predict API calls. Their model achieves a 90\% accuracy in the top-3 recommendations.

Nguyen \etal~\cite{Nguyen:icse2012} proposed GraPacc, a context-sensitive code completion model trained on a database of API usage patterns. These patterns are then matched to a given code under development to support code completion. GraPacc achieves up to 95\% precision and 92\% recall. A similar approach was later on proposed by Niu \etal~\cite{niu2017api} for API completion in Android: Given an API method as a query, their approach recommends a set of relevant API usage patterns. They report an 18\% improvement of F-Measure when comparing to pattern extraction using frequent-sequence mining. 

Tu \etal~\cite{Tu:fse2014} introduced a cache component to exploit the ``localness of code'' in the n-gram model. Results show that since the code is locally repetitive, localized information can be used to improve performance. The enhanced model outperforms standard n-gram models by up to 45\% in accuracy. In a related work, Franks\etal~\cite{franks2015cacheca} implemented \emph{CACHECA}, an Eclipse auto-completion plugin exploiting the aforementioned cache language model \cite{Tu:fse2014}. In comparison to Eclipse built-in suggestions, their tool improves the accuracy of top 1 and top 10 suggestions by 26\% and 34\%, respectively.

Nguyen \etal \cite{nguyen2015graph} presented GraLan, a graph-based statistical language model that the authors instantiated to recommend the next API element needed in a given code, where an API element is a method call together with the control units (\eg \texttt{if} statements) needed for its usage. The reported empirical evaluation showed that GraLan can correctly recommend the correct API element in 75\% of cases within the first five candidates.

Hou and Pletcher \cite{hou2011evaluation} evaluated three mechanisms to enhance code completion techniques, namely sorting, filtering, and grouping. Also this works focuses on code completion related to API methods and the outcome of their study is an assessment of the effectiveness of fourteen different configurations of the three mechanisms.

Asaduzzaman \etal \cite{Asaduzzaman:icsme2017} proposed a technique to recommend developers with examples of framework extensions. Given a class under development, the approach recommends code examples showing how to integrate frameworks in specific extension points. While the approach by Asaduzzaman \etal recommends relatively large code completion fragments, it is limited to a specific scenario, \ie framework extension points, whereas the approaches we experiment with are more general in that respect.

Hellendoorn and Devanbu~\cite{Hellendoorn:fse2017} proposed further improvements to the cached models aimed at considering specific characteristics of code (\eg unlimited, nested, and scoped vocabulary). Then, they compare their model with DL-based models, showing its superiority. Also, they show that the two families of techniques can be combined together, leading to an unprecedented 1.25 bits of entropy per token. Karampatsis \etal~\cite{Karampatsis:DLareBest}, a few years later, suggested instead that neural networks are the best language-agnostic algorithm for code completion. They proposed to overcome the \emph{out-of-vocabulary problem} by using \textit{Byte Pair Encoding} \cite{bpe}. In addition, the proposed neural network is able to dynamically adapt to different projects. Their best model outperforms n-gram models, achieving an entropy of 1.03 bits. 

Kim \etal~\cite{kim2020code} leveraged the Transformers neural network architecture for code completion. They provide the syntactic structure of code to the network by using information from the Abstract Syntax Tree to fortify the self-attention mechanism. Among the several models they experiment with, the best one reached a MRR up to 74.1\% in predicting the next token.

Alon \etal~\cite{alon2019structural} addressed the problem of code completion with a language agnostic approach named Structural Language Model. It leverages the syntax to model the code snippet as a tree. The model, based on LSTMs and Transformers, receives an AST representing a partial expression (statement), with some missing consecutive tokens to complete. Their best model reached state-of-the-art performance with an exact match accuracy for the top prediction of 18.04\%.

Svyatkovskiy \etal~\cite{svyatkovskiy2020intellicode} introduced IntelliCode Compose, a general-purpose multilingual code completion tool capable of predicting code sequences of arbitrary token types. They do not leverage high-level structural representation, such as AST, and use subtokens to overcome the \emph{out-of-vocabulary problem}. Their model can recommend an entire statement, and achieves a perplexity of 1.82 for the Python programming language.

Liu \etal~\cite{Liu:ase2020} presented a Transformer-based neural architecture pre-trained with the goal of incorporating both code understanding and generation tasks. Afterwards, the model was then fine-tuned on the classic code completion task (\ie predicting the next token to write).

A problem related to code completion has also been tackled by Watson \etal \cite{Watson:icse2020}: The authors exploit a sequence-to-sequence model to recommend assert statements for a given Java test case. This technique is able to generate a specific type of code statement, with a top-1 accuracy of 31\%. Also, Kanade \etal \cite{kanade2020} show how code embeddings can support code-related tasks, including  \emph{variable misuse and repair}, related to code completion when focusing on a single token.

Svyatkovskiy \etal \cite{svyatkovskiy2020fast} proposed a different perspective on neural code completion, shifting from a generative task to a learning-to-rank task. Their model is used to rerank the recommendations provided via static analysis, being cheaper in terms of memory footprint than generative models. To this aim, Avishkar \etal \cite{bhoopchand2016learning} proposed a neural language model for code suggestion in Python, aiming to capture long-range relationships among identifiers exploiting a sparse pointer network.

To address the out-of-vocabulary problem in standard neural language models, Jian \etal \cite{li2017code} proposed a pointer mixture deep learning model for Python benefiting from the pointer copy mechanism. Such architecture helps the model to generate an out-of-vocabulary word from local context through a pointer component when generating a within-vocabulary token is not possible.

A considerable step forward, has been taken recently by Aye and Kaiser \cite{aye2020sequence} proposing a novel language model to predict the next top-k tokens while taking into consideration some real-world constraints such as (i) prediction latency, (ii) size of the model and its memory footprint, and (iii) validity of suggestions.  
Chen \etal \cite{chen2020holistic} proposed a deep learning model for API recommendation combining structural and textual code information based on an API context graph and code token network. The evaluation model significantly outperforms the existing graph-based statistical approach and the tree-based deep learning approach for API recommendation.

To the best of our knowledge, our work is the first to present a comprehensive study on the effectiveness of Transformer models for code completion tasks, \textit{pushing this problem forward by attempting the automatic generation of an entire code block} (\eg the body of a {\tt for} statement). 
\subsection{Studies About the Effectiveness of Code Completion Approaches}

Although code completion techniques are likely to be beneficial for developers, their limitations (\eg prediction latency, accuracy) can bound their practical usefulness. For this reason, several studies investigated the effectiveness of code completion techniques.

Jin and Servant \cite{jin2018hidden} investigated the effect of different recommendation list lengths on the developers' productivity. They found that lengthy suggestion lists are not uncommon and reduce the developer's likelihood of selecting one of the recommendations. 

Lin \etal \cite{jiang2019machine} focus on the performance of a \emph{code2vec} \cite{alon2019code2vec} model, in the context of method name recommendation. The authors retrain the model on a different dataset and assess it in a more realistic setting where the training dataset does not contain any record from evaluation projects. The results suggest that while the dataset change had little impact on the model's accuracy, the new \emph{project-based} setting negatively impacted the model. Lin \etal \cite{jiang2019machine} also evaluated the usefulness of \emph{code2vec} suggestions by asking developers to assess the quality of suggestions for non-trivial method names. The evaluation results show the model rarely works when it is needed in practice. Further investigation also revealed that around half of successful recommendations (48\%) occur for simpler scenarios, such as setter/getter methods or when the recommended name is copied from the method body source code. 

Hellendoorn \etal \cite{HellendoornPGB19} studied 15,000 real code completions from 66 developers founding that typically-used code completion benchmarks --- \eg produced by artificially masking tokens --- may misrepresent actual code completion tasks. The study by Hellendoorn \etal suggests that further research is needed to assess the actual applicability of DL-based code completion to the real-world. This is however out of scope for our work, because our aim is to assess the capability of DL models to predict non-trivial portions of code going beyond a single method call or parameter.

Liu \etal \cite{liu2020deep} investigate the performance of deep learning-based approaches for generating code from requirement texts. For that, they assessed five state-of-the-art approaches on a larger and more diverse dataset of pairs of software requirement texts and their validated implementation as compared to those used in the literature.
The evaluation results suggest that the performance of such approaches, in terms of common metrics (\eg BLEU score), is significantly worse than what was reported in the literature. The authors attribute this observation to the relatively small datasets on which such models are evaluated. 

Similarly, Aye \etal \cite{aye2020learning} investigate the impact of using real-world code completion examples (\ie code completion acceptance events in the past) for training models instead of artificial examples sampled from code repositories. The usage of such realistic data on n-gram and transformer models suggests a significant accuracy decrease. Later, an A/B test conducted with Facebook developers confirmed that the autocompletion usage increases by around 6\% for models trained on real-world code completion examples.

Our work, differently from previous studies, aims at assessing the capability of state-of-the-art Transformer-based models in predicting non-trivial snippets of code. In contrast, it is out of this study scope to assess the developer's perception of the prediction models that would require an extensive study with developers.

%% file: conclusion.tex
% !TEX root = main.tex
%%%%%%%%%%%%%%%%%%%%%%%%%%%%%%%%%%%%%%%%
%%%%%%%%%%%%%%%%%%%%%%%%%%%%%%%%%%%%%%%%
\section{Conclusion} \label{sec:conclusion}
%%%%%%%%%%%%%%%%%%%%%%%%%%%%%%%%%%%%%%%%
%%%%%%%%%%%%%%%%%%%%%%%%%%%%%%%%%%%%%%%%
We investigated the ability of Transformer-based DL-models in dealing with code completion tasks having a different level of difficulty, going from the prediction of a few tokens within the same code statement, up to the entire code blocks we masked. Among the three models we experimented with, namely T5 \cite{raffel2019exploring}, \roberta \cite{Delvin:2019}, and the cached \textit{n}-gram model \cite{Hellendoorn:fse2017}, the T5 resulted to be the most effective in supporting code completion. 

Our study provided a series of highlights that will guide our future research. First, when the code to complete spans over multiple statements (two in the case of our experiments), these models, with the training we performed, are still far from being a valuable solution for software developers. Indeed, even the best-performing model (T5) struggles in guessing entire code blocks. However, the performance we reported should not be seen as an ``upper bound'' for these techniques, since larger models may be trained on more data can be adopted (\eg the recently proposed GitHub Copilot \cite{copilot}) and different training strategies could help in achieving better results (\eg Tufano \etal \cite{tufano:testGeneration} showed that pre-training on English text helps transformer models in improving performance even in code-related tasks). Besides working on these research directions we also plan to investigate alternative solutions mixing, for example, retrieval-based and DL-based solutions. 

Second, the confidence of the predictions generated by the T5 turned out to be a very reliable proxy for the quality of its predictions. This is something fundamental for building tools around this model, as it can be used by developers to just ignore low-confidence recommendations. Future studies will investigate how the developers perceive the usefulness of recommendations having different characteristics, including length, confidence, and covered code constructs.

Finally, a user study is also needed to understand what is the level of accuracy (in terms of perfect predictions) needed to consider tools built around these models as effective for developers. In other words, it is important to understand the ``percentage of wrong predictions'' a developer can accept before considering the tool counterproductive. Such a study is also part of our research agenda. 

%% file: bios.tex
% !TEX root = main.tex
\begin{IEEEbiography}[{\includegraphics[width=1in,height=1.25in,clip,keepaspectratio]{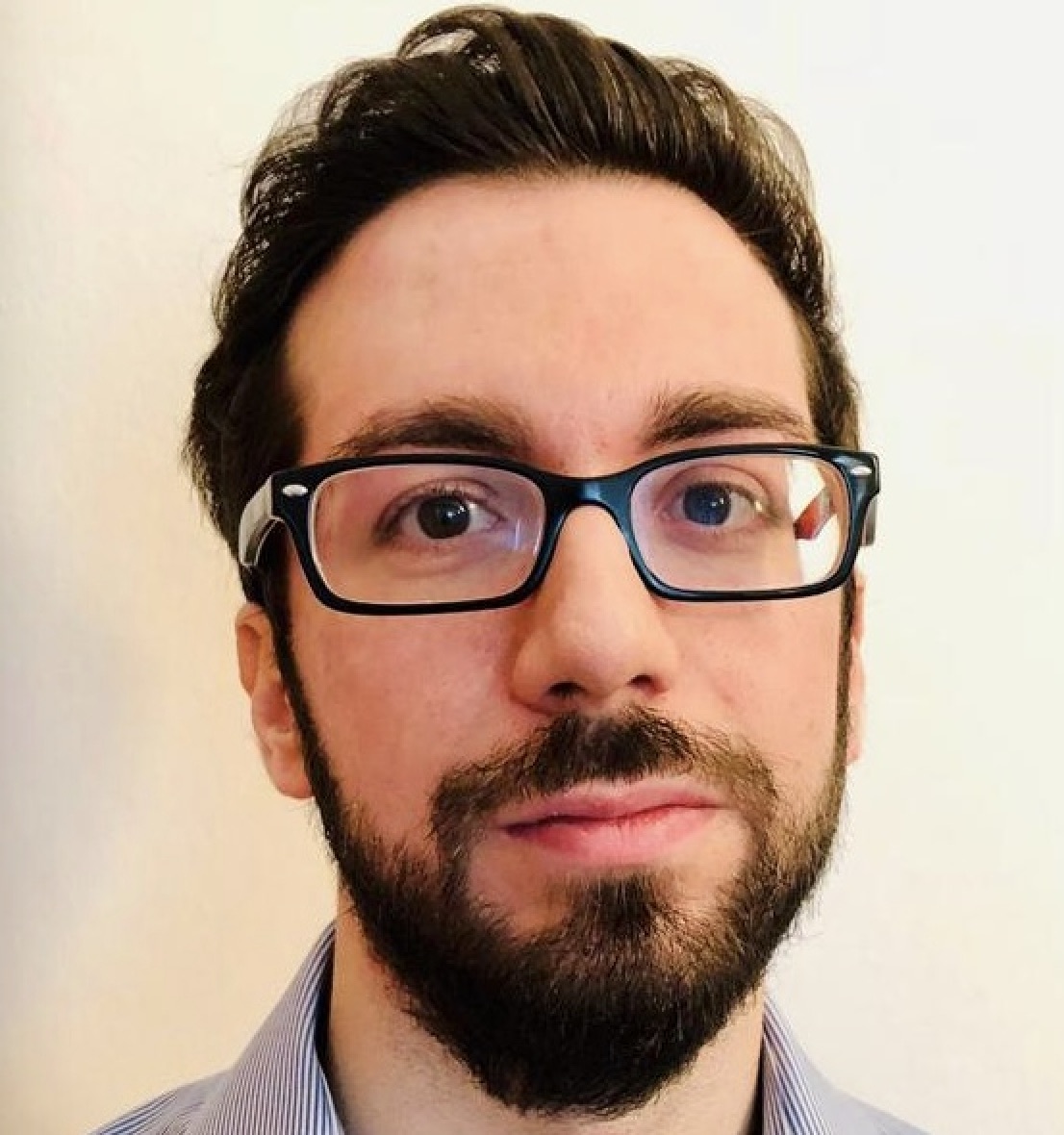}}]{Matteo Ciniselli} is a Ph.D. student in the Faculty of Informatics at the Universit\`a della Svizzera italiana (USI), Switzerland, where he is part of the Software Institute. He received his MSc. in Mathematical Engineering from Politecnico di Milano, Italy, in April 2015. His research interests include the study of deep-learning models to support code-related tasks. More information available at: \url{https://www.inf.usi.ch/phd/cinism}.
\end{IEEEbiography}

\begin{IEEEbiography}[{\includegraphics[width=1in,height=1.25in,clip,keepaspectratio]{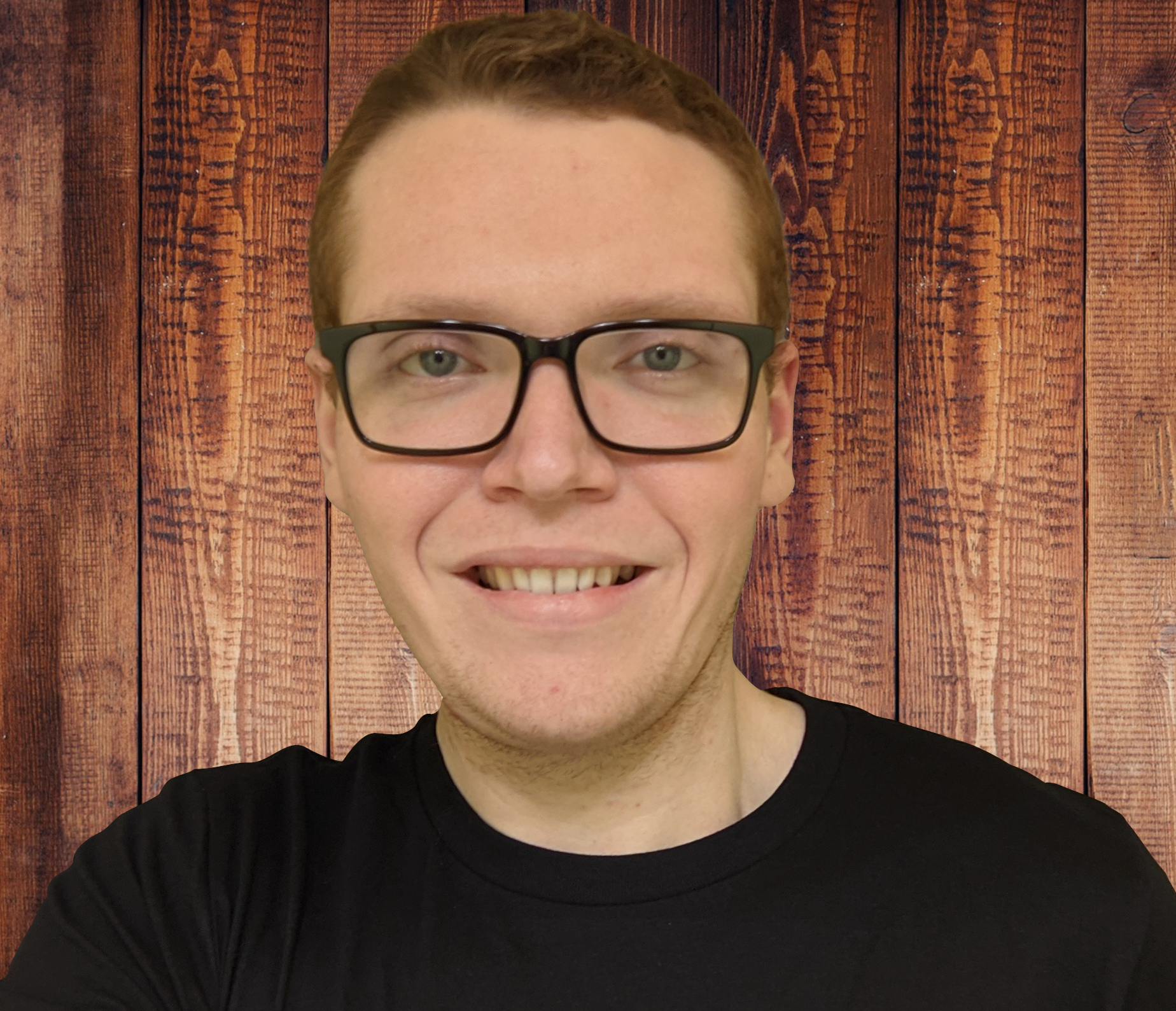}}]{Nathan Cooper} received a B.S. degree in Software Engineering from the University of West Florida in 2018. He is currently a Ph.D. candidate in Computer Science at William \& Mary under the advisement of Dr. Denys Poshyvanyk and is a member of the Semeru Research group. He has research interests in Software Engineering, Machine / Deep Learning applications for Software Engineering, information retrieval, and question \& answering applications for Software Engineering. He has published in the top peer-reviewed Software Engineering venues ICSE and MSR. He has also received the ACM SIGSOFT Distinguished paper award at ICSE'20. More information is available at \url{https://nathancooper.io/#/}.
\end{IEEEbiography}

\begin{IEEEbiography}[{\includegraphics[width=1in,height=1.25in,clip,keepaspectratio]{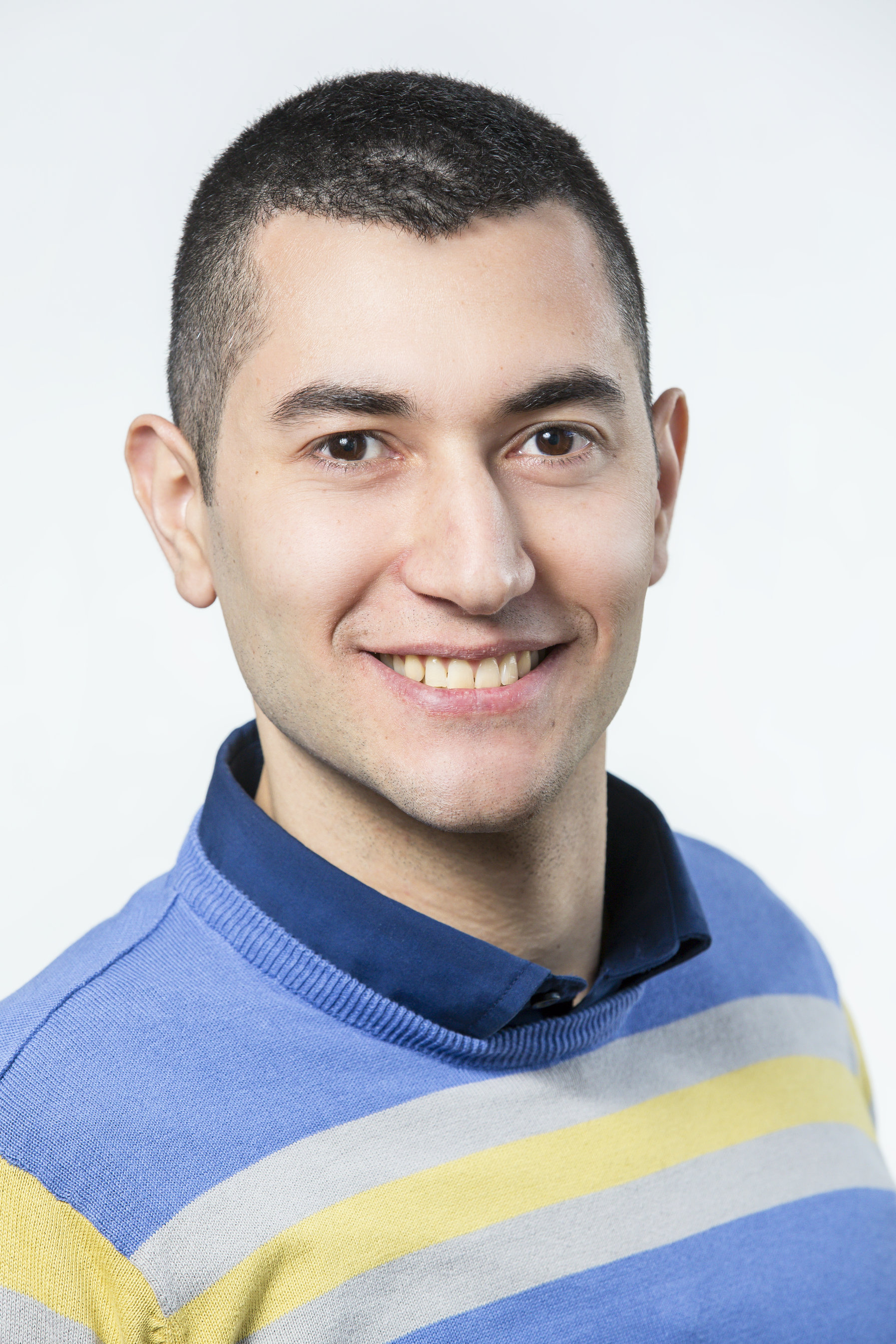}}]{Luca Pascarella} is a postdoctoral researcher at the Universit\`a della Svizzera italiana (USI), Switzerland, where he is part of the Software Institute. He received his Ph.D. in Computer Science from the Delft University of Technology (TU Delft), The Netherlands, in 2020. His broader mission aims to smooth engineering tasks through data-driven algorithms, which leverage the large amount of information recorded during modern engineering processes. His research interests include empirical software engineering, mining software repository, and code review. He received an ACM SIGSOFT Distinguished Paper Award at MSR 2017 and a Best Paper Award Honorable Mention at CSCW2018. More information available at: \url{https://lucapascarella.com}.
\end{IEEEbiography}

\begin{IEEEbiography}[{\includegraphics[width=1in,height=1.25in,clip,keepaspectratio]{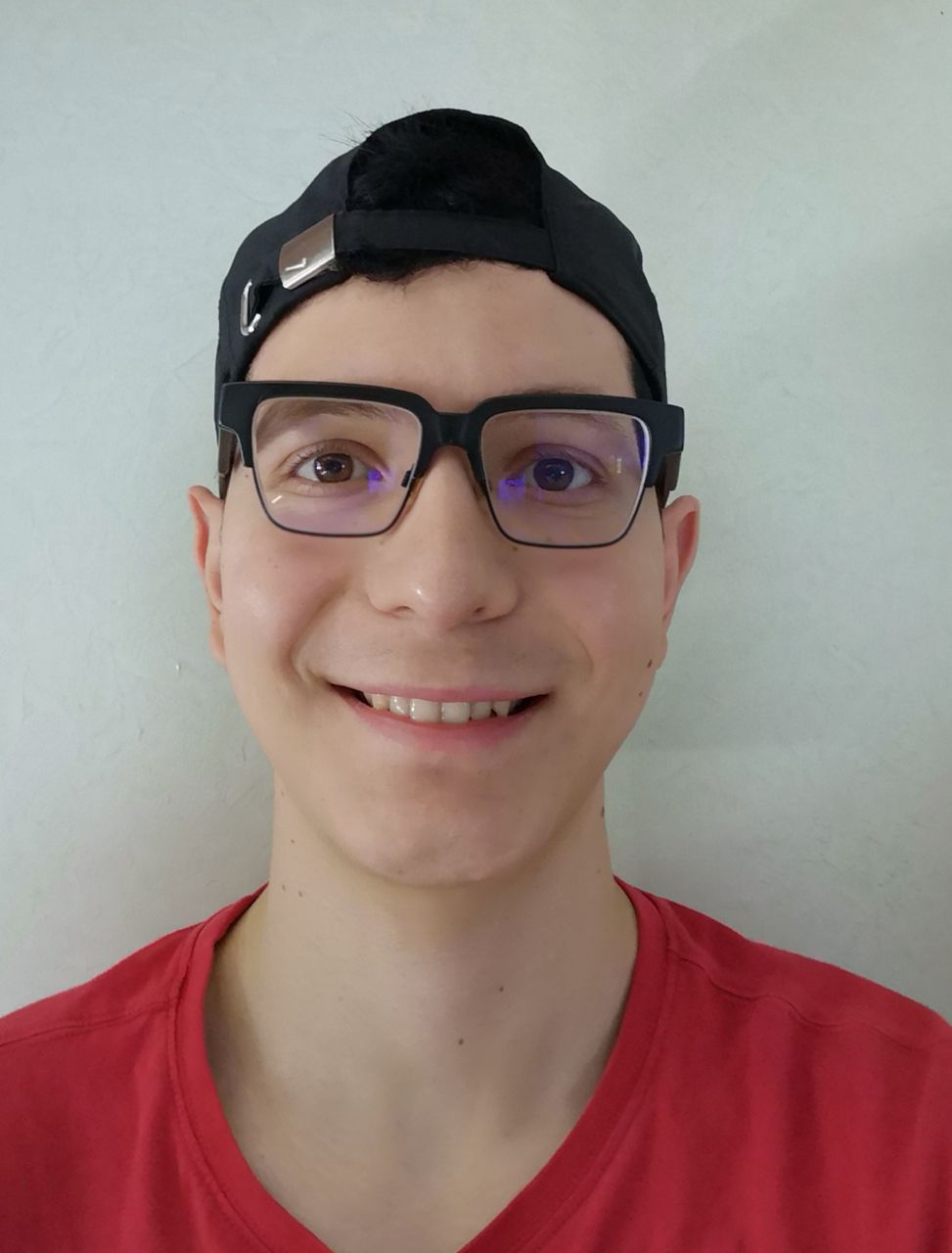}}]{Antonio Mastropaolo} is a Ph.D. student in the Faculty of Informatics at the Universit\`a della Svizzera italiana (USI), Switzerland, where he is part of the Software Institute. He received his MSc. in Software System Security from Universit\`a degli studi del Molise, Italy, in July 2020. His research interests include the study and the application of deep-learning techniques to foster code-related tasks. More information available at: \url{https://antoniomastropaolo.com}.
\end{IEEEbiography}

\begin{IEEEbiography}[{\includegraphics[width=1in,height=1.25in,clip,keepaspectratio]{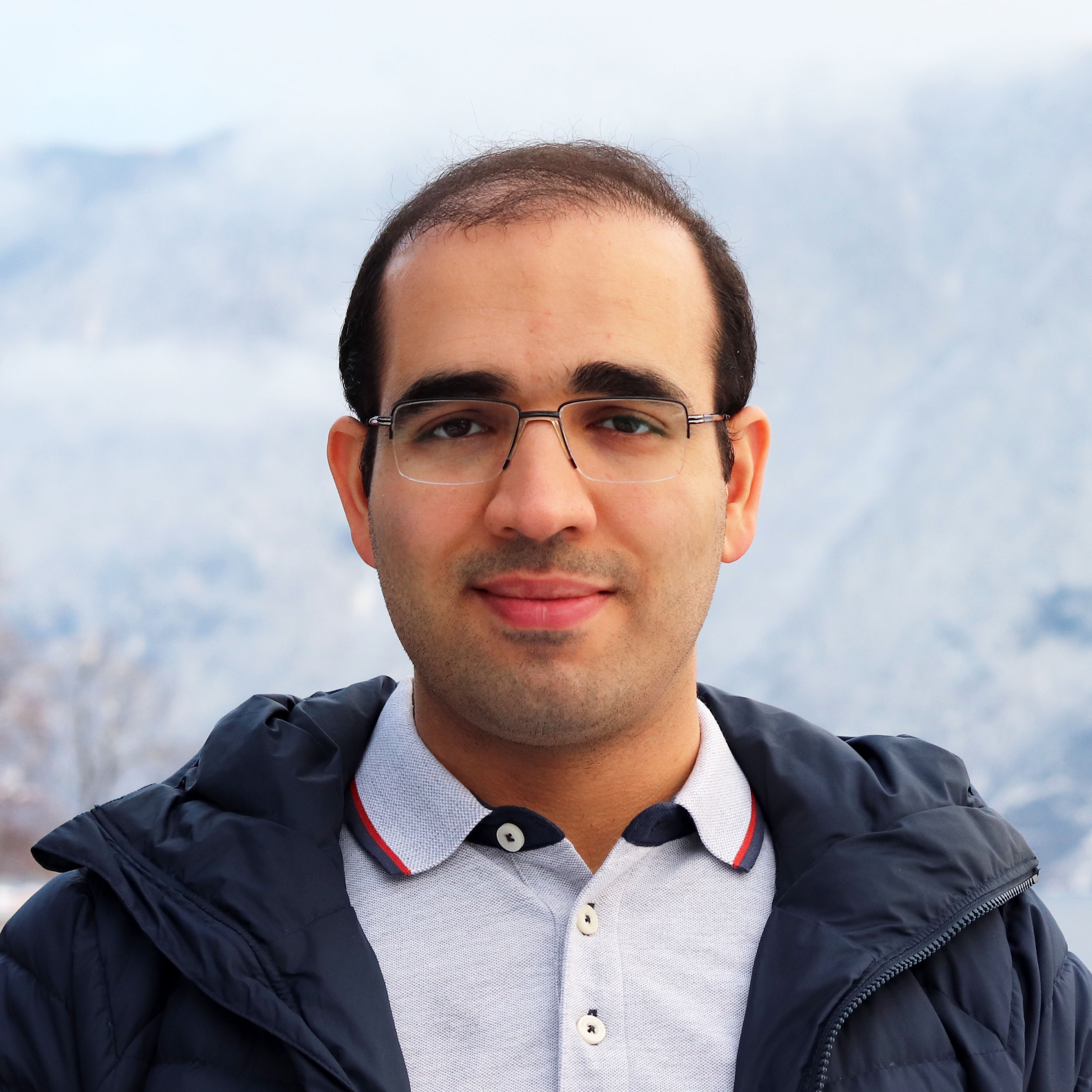}}]{Emad Aghajani} is a postdoctoral researcher in the SEART research group, at Software Institute, the Universit\`a della Svizzera italiana (USI), Switzerland. He finished his Ph.D. studies in Computer Science in 2020 under the supervision of Prof. Michele Lanza and Prof. Gabriele Bavota at USI. He received his M.Sc. in Software Engineering from Sharif University of Technology, Iran, in 2016. His research interests include software evolution, software maintenance, mining software repositories, and empirical software engineering. More information is available at: \url{https://emadpres.github.io/}.
\end{IEEEbiography}

\begin{IEEEbiography}[{\includegraphics[width=1in,height=1.25in,clip,keepaspectratio]{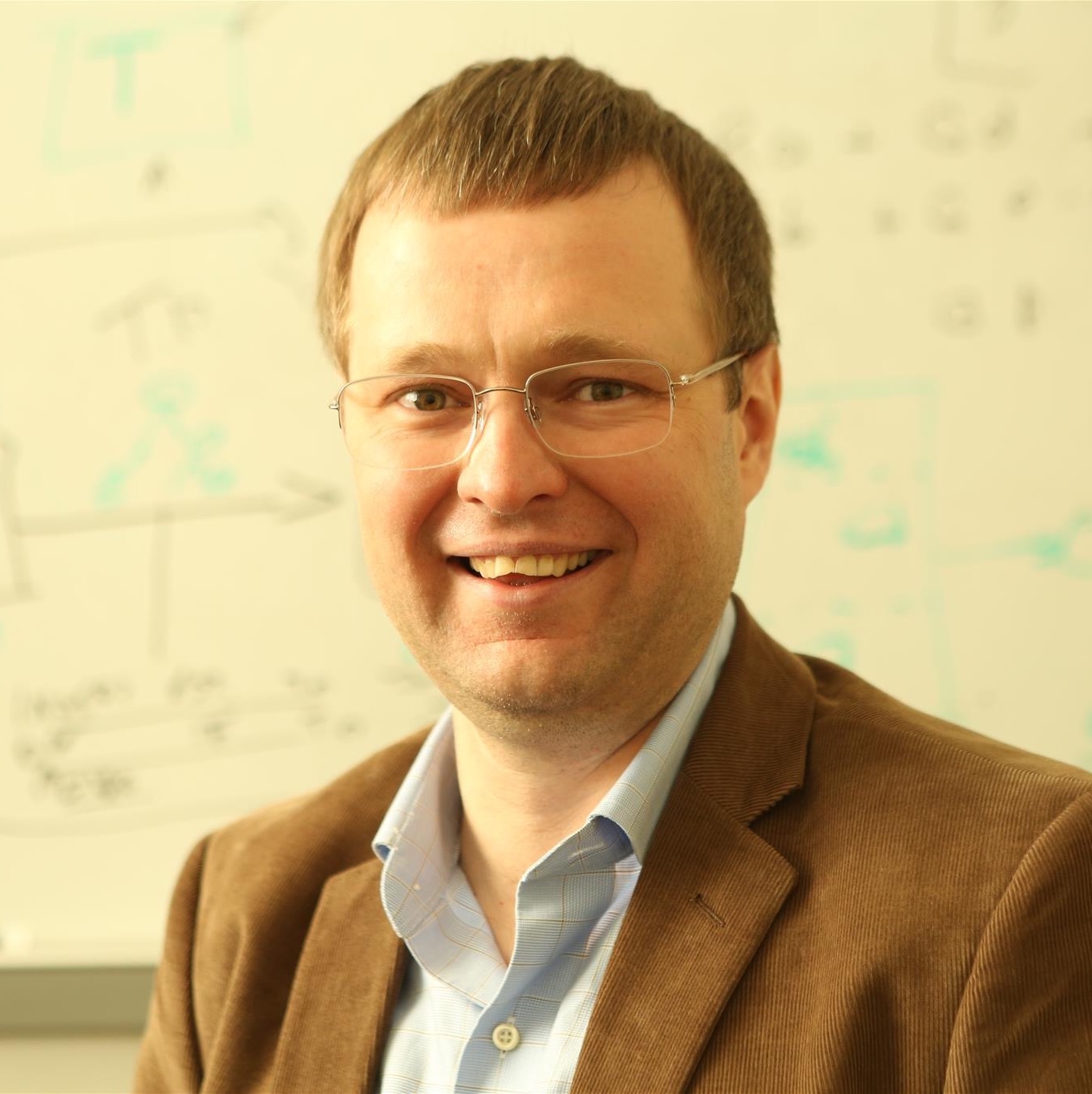}}]{Denys Poshyvanyk} is a Professor of Computer Science at William and Mary. He received the MS and MA degrees in Computer Science from the National University of Kyiv-Mohyla Academy, Ukraine, and Wayne State University in 2003 and 2006, respectively. He received the PhD degree in Computer Science from Wayne State University in 2008. He served as a program co-chair for ASE'21, MobileSoft'19, ICSME'16, ICPC'13, WCRE'12 and WCRE'11. He currently serves on the editorial board of IEEE Transactions on Software Engineering (TSE), ACM Transactions on Software Engineering and Methodology (TOSEM), Empirical Software Engineering Journal (EMSE, Springer), Journal of Software: Evolution and Process (JSEP, Wiley) and Science of Computer Programming. His research interests include software engineering, software maintenance and evolution, program comprehension, reverse engineering and software repository mining. His research papers received several Best Paper Awards at ICPC'06, ICPC'07, ICSM'10, SCAM'10, ICSM'13, CODAPSY'19 and ACM SIGSOFT Distinguished Paper Awards at ASE'13, ICSE'15, ESEC/FSE'15, ICPC'16, ASE'17, ESEC/FSE'19 and ICSE'20. He also received the Most Influential Paper Awards at ICSME'16, ICPC'17 and ICPC'20. He is a recipient of the NSF CAREER award (2013).  He is a member of the IEEE and ACM. More information is available at: \url{http://www.cs.wm.edu/~denys/}.
\end{IEEEbiography}

\begin{IEEEbiography}[{\includegraphics[width=1in,height=1.25in,clip,keepaspectratio]{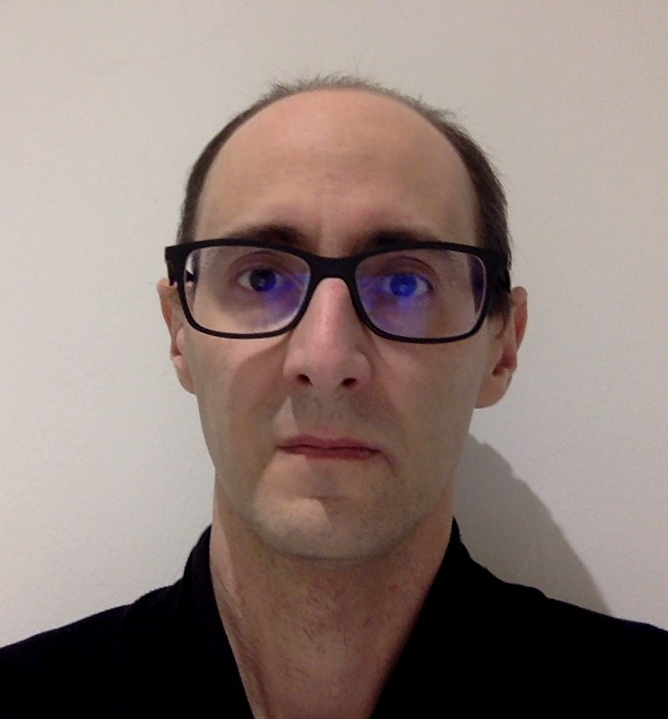}}]{Massimiliano Di Penta} is a full professor at the University of Sannio, Italy. His research interests include software maintenance and evolution, mining software repositories, empirical software engineering, search-based software engineering, and service-centric software engineering. He is an author of over 300 papers appeared in international journals, conferences, and workshops. He serves and has served in the organizing and program committees of more than 100 conferences, including ICSE, FSE, ASE, ICSME. He is editor-in-chief of the Journal of Software: Evolution and Processes edited by Wiley. He is in the editorial board of ACM Transactions on Software Engineering and Methodology, and of the Empirical Software Engineering Journal. He has served the editorial board of the IEEE Transactions on Software Engineering.
\end{IEEEbiography}

\begin{IEEEbiography}[{\includegraphics[width=1in,height=1.25in,clip,keepaspectratio]{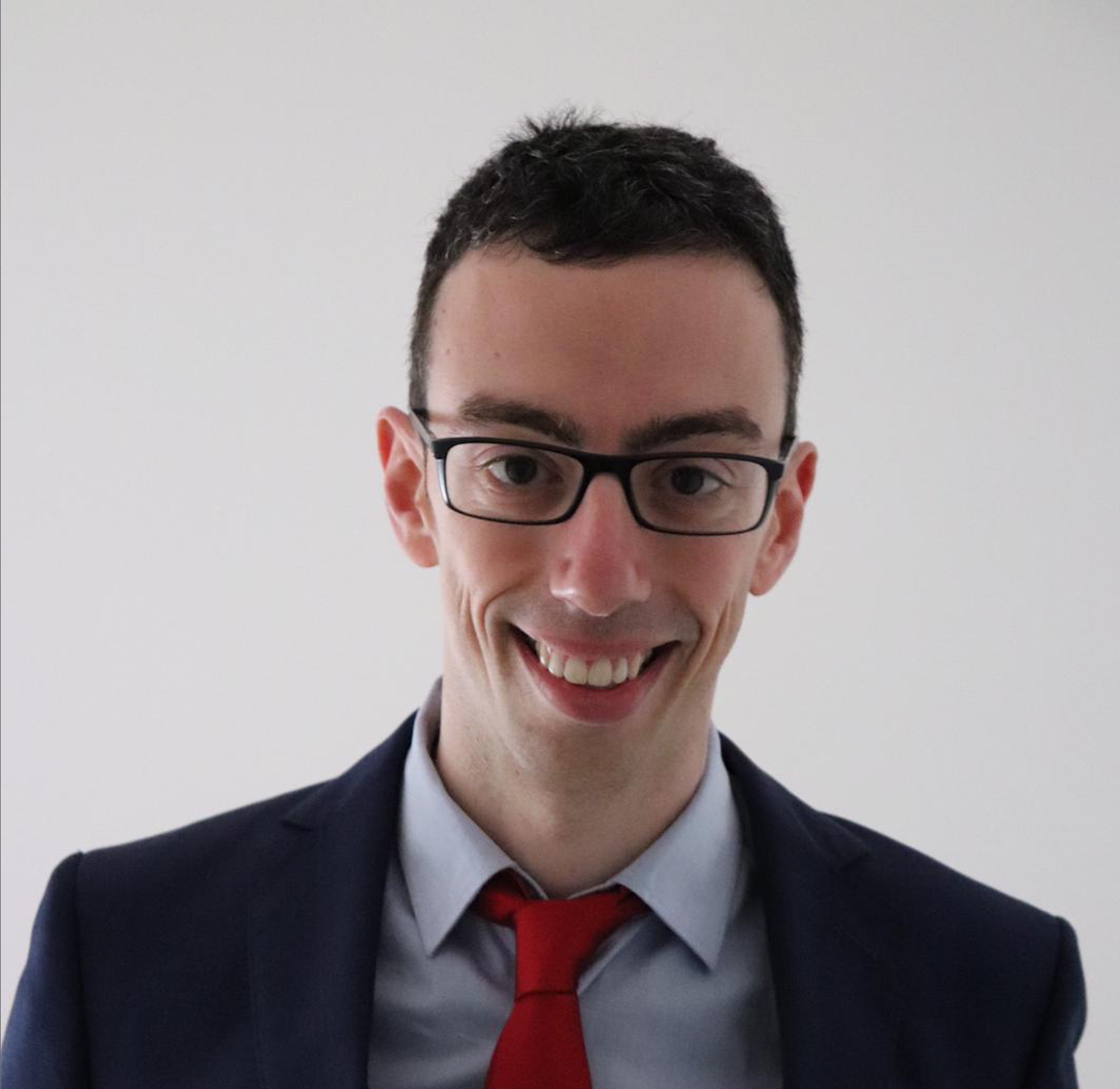}}]{Gabriele Bavota} Gabriele Bavota is an associate professor at the Faculty of Informatics of the Universit\`a della Svizzera italiana (USI), Switzerland, where he is part of the Software Institute and he leads the SEART research group. He received the PhD in Computer Science from the University of Salerno, Italy, in 2013. His research interests include software maintenance and evolution, code quality, mining software repositories, and empirical software engineering. On these topics, he authored over 140 papers appeared in international journals and conferences and has received four ACM Sigsoft Distinguished Paper awards at the three top software engineering conferences: ASE 2013 and 2017, ESEC-FSE 2015, and ICSE 2015. He also received the best/distinguished paper award at SCAM 2012, ICSME 2018, MSR 2019, and ICPC 2020.
He is the recipient of the 2018 ACM Sigsoft Early Career Researcher Award for outstanding contributions in the area of software engineering as an early career investigator and the principal investigator of the DEVINTA ERC project. More information is available at: \url{https://www.inf.usi.ch/faculty/bavota/}.
\end{IEEEbiography}